\documentclass[runningheads]{llncs}
\usepackage{booktabs}

\usepackage{hyperref}
\usepackage{stmaryrd}
\usepackage{amsmath}
\usepackage{amssymb}
\usepackage[square,comma,sort]{natbib}   
\usepackage{ntheorem}
\usepackage{mdwlist}
\usepackage{subcaption}
\usepackage{enumitem}
\usepackage{wrapfig}
\usepackage{tikz}
\usepackage{tkz-berge}
\DeclareMathOperator{\argmax}{argmax}

\begin{document}

\newcommand{\shortcite}[1]{[\citeyear{#1}]}
\def\cite{\citep}

\newcommand*{\QED}{\hfill\ensuremath{\square}}%
\newenvironment{rtheorem}[1]{\medskip\noindent\textbf{Theorem~\ref{#1}.}\begin{itshape}}{\end{itshape}}
\newenvironment{rproposition}[1]{\medskip\noindent\textbf{Proposition~\ref{#1}.}\begin{itshape}}{\end{itshape}}
\newenvironment{rlemma}[1]{\medskip\noindent\textbf{Lemma~\ref{#1}.}\begin{itshape}}{\end{itshape}}
\newenvironment{rconjecture}[1]{\medskip\noindent\textbf{Conjecture~\ref{#1}.}\begin{itshape}}{\end{itshape}}


\hbadness=10000
\vbadness=10000
\newcommand{\newpar}[1]{\paragraph{#1.}}

\newcommand{\coursename}{COST Summer school on Social Choice}

\newcommand{\handout}[5]{
   \renewcommand{\thepage}{#1 \arabic{page}}
   \noindent
   \begin{center}
   \framebox{
      \vbox{
    \hbox to 5.78in { {\bf \coursename}
         \hfill #2 }
       \vspace{4mm}
       \hbox to 5.78in { {\Large \hfill #5  \hfill} }
       \vspace{2mm}
       \hbox to 5.78in { {\it #3 \hfill #4} }
      }
   }
   \end{center}
   \vspace*{4mm}
}
\newcommand{\labeq}[2]{
\begin{equation}
\label{eq:#1} #2
\end{equation}}
\newcommand{\lecture}[5]{\handout{#1}{#3}{Lecturer:
#4}{Scribe: #5}{Lecture #1: #2}}

\newcommand{\exam}[1]{\handout{#1}{}{}{}{Exam}}
\newcommand{\homeexam}[2]{\handout{#1}{}{}{Due:
#2}{Home Exam}}
\newcommand{\examanswer}[1]{\handout{X}{}{by: #1}{}{Exam - Answers}}

\newcommand{\problemset}[2]{\handout{#1}{}{}{Due:
#2}{Problem Set #1}}


\newcommand{\exanswer}[2]{\handout{#1}{}{by: #2}{}{Problem Set #1 - Answers}}

\def\range{\text{range}}
%

%

\newcommand{\dis}{\mathop{\mbox{\rm d}}\nolimits}
\newcommand{\per}{\mathop{\mbox{\rm per}}\nolimits}
\newcommand{\area}{\mathop{\mbox{\rm area}}\nolimits}
\newcommand{\ccw}{\mathop{\rm ccw}\nolimits}
\newcommand{\DIST}{\mathop{\mbox{\rm DIST}}\nolimits}
\newcommand{\OP}{\mathop{\mbox{\it OP}}\nolimits}
\newcommand{\OPprime}{\mathop{\mbox{\it OP}^{\,\prime}}\nolimits}
\newcommand{\ihat}{\hat{\imath}}
\newcommand{\jhat}{\hat{\jmath}}
\newcommand{\abs}[1]{\mathify{\left| #1 \right|}}
\newcommand{\ceil}[1]{\left\lceil #1\right\rceil}
\newcommand{\floor}[1]{\left\lfloor #1\right\rfloor}
\newcommand{\hatv}[1]{\hat{\vec{#1}}}
\newcommand{\step}[1]{\stackrel{#1}{\rightarrow}}
\newenvironment{proof-sketch}{\noindent{\bf Sketch of Proof}\hspace*{1em}}{\qed\bigskip}
\newenvironment{proof-idea}{\noindent{\bf Proof Idea}\hspace*{1em}}{\qed\bigskip}
\newenvironment{proof-of-lemma}[1]{\noindent{\bf Proof of Lemma #1}\hspace*{1em}}{\qed\bigskip}
\newenvironment{proof-attempt}{\noindent{\bf Proof Attempt}\hspace*{1em}}{\qed\bigskip}
\newenvironment{proofof}[1]{\noindent{\bf Proof}
of #1:\hspace*{1em}}{\qed\bigskip}
\def\ol{\overline}

\def\top{\text{top}}

\def\defeq{\triangleq}
\newcommand{\FOR}{{\bf for}}
\newcommand{\TO}{{\bf to}}
\newcommand{\DO}{{\bf do}}
\newcommand{\WHILE}{{\bf while}}
\newcommand{\AND}{{\bf and}}
\newcommand{\IF}{{\bf if}}
\newcommand{\THEN}{{\bf then}}
\newcommand{\ELSE}{{\bf else}}

\makeatletter
\def\fnum@figure{{\bf Figure \thefigure}}
\def\fnum@table{{\bf Table \thetable}}
\long\def\@mycaption#1[#2]#3{\addcontentsline{\csname
  ext@#1\endcsname}{#1}{\protect\numberline{\csname
  the#1\endcsname}{\ignorespaces #2}}\par
  \begingroup
    \@parboxrestore
    \small
    \@makecaption{\csname fnum@#1\endcsname}{\ignorespaces #3}\par
  \endgroup}
\def\mycaption{\refstepcounter\@captype \@dblarg{\@mycaption\@captype}}
\makeatother

\newcommand{\figcaption}[1]{\mycaption[]{#1}}
\newcommand{\tabcaption}[1]{\mycaption[]{#1}}
\newcommand{\head}[1]{\chapter[Lecture \##1]{}}
\newcommand{\mathify}[1]{\ifmmode{#1}\else\mbox{$#1$}\fi}
\newcommand{\bigO}O
\newcommand{\set}[1]{\mathify{\left\{ #1 \right\}}}
\newcommand\card[1]{\left| #1 \right|}
\newcommand\cardi[1]{| #1 |}
\newcommand\tup[1]{\left\langle #1 \right\rangle}
\newcommand\sett[2]{\left\{ \left. #1 \;\right\vert #2 \right\}}
\newcommand\settright[2]{\left\{ #1 \;\left\vert\; #2 \right.\right\}}

\def\half{\frac{1}{2}}
\def\trd{\frac{1}{3}}
\def\ttrd{\frac{2}{3}}
\def\qd{\frac{1}{4}}
\def\sx{\frac{1}{6}}

\def\vecL{\boldsymbol{L}}

\newcommand{\enc}{{\sf Enc}}
\newcommand{\dec}{{\sf Dec}}
\newcommand{\E}{\mathbb{E}}
\newcommand{\Var}{{\rm Var}}
\newcommand{\Z}{{\mathbb Z}}
\newcommand{\F}{{\mathbb F}}
\newcommand{\integers}{{\mathbb Z}^{\geq 0}}
\newcommand{\R}{{\mathbb R}}
\newcommand{\Q}{{\cal Q}}
\newcommand{\eqdef}{{\stackrel{\rm def}{=}}}
\newcommand{\from}{{\leftarrow}}
\newcommand{\vol}{{\rm Vol}}
\newcommand{\poly}{{\rm poly}}
\newcommand{\ip}[1]{{\langle #1 \rangle}}
\newcommand{\wt}{{\rm wt}}
\renewcommand{\vec}[1]{{\boldsymbol #1}}
\newcommand{\mspan}{{\rm span}}
\newcommand{\rs}{{\rm RS}}
\newcommand{\RM}{{\rm RM}}
\newcommand{\Had}{{\rm Had}}
\newcommand{\calc}{{\cal C}}
\def\calD{{\cal D}}
\def\calO{{\cal O}}
\def\calT{{\calT}}
\def\eps{\varepsilon}
\newcommand{\ind}[1]{\left\llbracket #1 \right\rrbracket}

\newcommand{\fig}[4]{
        \begin{figure}
        \setlength{\epsfysize}{#2}
        \vspace{3mm}
        \centerline{\epsfbox{#4}}
        \caption{#3} \label{#1}
        \end{figure}
        }

\newcommand{\ord}{{\rm ord}}

\providecommand{\norm}[1]{\left\| #1 \right\|}
\newcommand{\embed}{{\rm Embed}}
\newcommand{\qembed}{\mbox{$q$-Embed}}
\newcommand{\calh}{{\cal H}}

\newcommand{\lp}{{\rm LP}}
\newcommand{\remove}[1]{{}}

\def\calR{{\cal R}}
\def\calU{{\cal U}}
\def\calL{{\cal L}}
\def\calX{{\cal X}}
\def\calW{{\cal W}}
\def\calA{{\cal A}}
\def\calB{{\cal B}}
\def\calJ{{\cal J}}
\newenvironment{sproof}{\noindent\textit{Proof sketch.}}{\medskip}

\newcommand{\rmr}[1]{}
\def\ol{\overline}
\def\ul{\underline}

\renewtheorem{conjecture}[theorem]{Conjecture}
\renewtheorem{proposition}[theorem]{Proposition}
\renewtheorem{lemma}[theorem]{Lemma}
\newtheorem{observation}[theorem]{Observation}
\renewtheorem{corollary}[theorem]{Corollary}
\def\({\left(}
\def\){\right)}

\usetikzlibrary{arrows,%
               petri,%
               topaths}%
							
							\def\cal{\mathcal}

\title{Strategyproof Facility Location for Three Agents on a Circle}
\author{Reshef Meir} 
\institute{Technion---Israel Institute of Technology\\
\email{reshefm@ie.technion.ac.il}}
\authorrunning{R. Meir}

%
\maketitle
\begin{abstract}
We consider the facility location problem in a metric space, focusing on the case of three agents. 
We show that selecting the reported location of each agent with probability proportional to the distance between the other two agents results in a mechanism that is strategyproof in expectation, and dominates the random dictator mechanism in terms of utilitarian social welfare. We further improve the upper bound for three agents on a circle to $\frac76$ (whereas random dictator obtains $\frac43$); and provide the first lower bounds for randomized strategyproof facility location in any metric space, using linear programming.  
\end{abstract}

\section{Introduction}
In a facility location problem, a central authority faces a set of agents who report their locations in some space, and needs to decide where to place a facility. It is typically assumed that each agent $i$ wants the facility to be placed as close as possible to her own location $a_i$. The challenge is to design a \emph{strategyproof} mechanism, such that reporting the truthful location is a weakly dominant strategy for every agent.  The designer may have additional goals, where the most common one is to minimize the utilitarian social cost---the sum of distances to agents' locations.

Strategyproof facility location mechanisms have been studied at least since the mid-20th century~\cite{black1948rationale}. In 2009, the agenda of approximation mechanisms without money was made  explicit in a paper by Procaccia and Tennenholtz~\shortcite{procaccia2009approximate,procaccia2013approximate}, who used facility location as their primary domain of demonstration due to its simplicity. Moreover, facility location is often a bridge between mechanism design and social choice~\cite{caragiannis2010approximation,MPR:2012:AIJ,feldman2016voting} and has applications to transport~\cite{moujahed2006reactive}, disaster relief~\cite{florez2015decision} and more.  Facility location is thus often used as a testbed for ideas and techniques in mechanism design and noncooperative multiagent systems. 

Most problems that include a single facility are by now well understood. For example, all deterministic strategyproof mechanisms on continuous and on discrete lines have been characterized~\cite{SV04,DFMN:2012:EC}, and it is well known that selecting the median agent location is both strategyproof and optimal in terms of utilitarian social cost~\cite{Moul80,procaccia2009approximate}. One strand of the literature seeks to characterize domains where median-like mechanisms exist~\cite{KM77,Nehring2007}.

For other domains, e.g. graphs that contain cycles, research following \cite{procaccia2009approximate} has focused on the minimal social cost that can be guaranteed by strategyproof mechanisms. For deterministic mechanisms even the existence of a single cycle in a graph entails that any strategyproof mechanism must be dictatorial on a subdomain, and thus has an approximation ratio that increases linearly with the number of agents $n$~\cite{SV04,DFMN:2012:EC}. Many variations of the problem have since been explored in the AI and multiagent systems community, including multiple facilities~\cite{escoffier2011strategy,serafino2015truthful,anastasiadis2018heterogeneous},  complex incentives and forms of strategic behavior~\cite{todo2011false,zou2015facility,sui2015approximately,filos2017facility}, and alternative design goals~\cite{alon2010strategyproof,feldman2013strategyproof,mei2016strategy}. 
The circle in particular has received much attention in the facility location literature~\cite{SV04,alon2010walking,alon2010strategyproof,DFMN:2012:EC,cai2016facility}, both because it is the simplest graph for which median-like mechanisms cannot work, and because it is an abstraction of actual problems like selecting a time-of-the-day or a server in a ring of computers. 

Yet, the fundamental strategyproof facility location problem for \emph{randomized mechanisms} remains almost unscathed. It is easy to show that the \emph{random dictator} (RD) mechanism obtains an approximation ratio of $2-\frac2n$ for any metric space~\cite{alon2010strategyproof,MPR:2012:AIJ}, and of course that $1$ is a lower bound. However except for lines and trees (where the deterministic Median mechanism is optimal), nothing else is known. 

To the best of our knowledge, the literature does not mention mechanisms that approximate the optimal social cost better than RD even for specific spaces like the circle, nor is there any lower bound higher than $1$.\footnote{Alon et al.~\shortcite{alon2010strategyproof} proposed a randomized strategyproof mechanism specifically for circles, called the \emph{hybrid mechanism}. They showed that it obtains the best possible approximation ratio for the \emph{minimax cost}, yet for the social cost it achieves a poor approximation ratio of $\frac{n-1}{2}$.}  The current paper focuses on narrowing this gap by proving tighter upper and lower bounds for three agents.

A variant of the problem on which there was more (negative) progress is when we allow arbitrary constraints on the location of the facility (e.g., where agents can be placed anywhere on a graph, but only 5 vertices are valid locations for the facility).
In the constrained variant, the RD mechanism obtains $3-\frac2n$ approximation and this is known to be tight for all strategyproof mechanisms. The upper bound holds for any metric space~\cite{MPR:2012:AIJ}, whereas the lower bound requires specific constructions on the $n$-dimensional binary cube~\cite{MAMR:2011:AAMAS,feldman2016voting}.  Anshelevich and Postl~\shortcite{anshelevich2017randomized} show a smooth transition of the RD approximation ratio from  $2-\frac2n$ to $3-\frac2n$ as the location of the facility becomes more constrained.  See \cite{SV_book} Section 5.3 for an overview of approximation results for a single facility.

\rmr{SP approximations of range voting 3 agents}

\subsection{Contribution}
Our main contribution is the introduction of two randomized mechanisms that beat the random dictator (RD) mechanism on a circle: the \emph{Proportional Circle Distance} (PCD) mechanism, which selects each reported location $a_i$ with probability proportional to the length $L_i$ of the arc facing agent~$i$; and the $q$-Quadratic Circle Distance mechanism ($q$-QCD) where the probability of selecting $a_i$ is proportional to $(\max\{(L_i)^2,q^2\})$.

We prove that PCD is strategyproof for any odd number of agents. For 3 agents, we show that PCD obtains an approximation ratio of $\frac54$ on the circle (in contrast to $\frac43$ by RD), and has a natural extension that is strategyproof and weakly dominates RD on any metric space. 
The $\frac14$-QCD mechanism is also strategyproof for 3 agents, and obtains an approximation ratio of $\frac76$ on the circle. 

For any  finite graph with $m$ vertices, there is a linear program of polynomial size that can compute the optimal randomized strategyproof mechanism. We use such programs to obtain first (but non-tight) lower bounds on the approximation ratio of any strategyproof mechanism on circles and on general graphs. See Table~\ref{tab:results_r} for a summary.


Some of our proofs use a combination of formal analysis and computer optimization. 
All full proofs appear in the appendix. Appendix~\ref{sec:plane} contains an analysis of the multi-dimensional median mechanism for 3 agents on the plane. A recent working paper by Goel and Hann-Caruthers~\shortcite{goel2019coordinate} solves a more general problem. 

\section{Preliminaries}
A domain of facility location problems is given by  $\tup{\calX,d}$, where $\calX$ is a set, and $d:\calX\times\calX \rightarrow \mathbb{R}_+$ is a distance metric. In this paper, $\calX$ is a (discrete or continuous) graph, and $d(x,y)$ is the length of the shortest path between $x$ and $y$. 
An \emph{instance} in the domain $\tup{\calX,d}$ is given by a profile $\vec a\in \calX^n$, where $n$ is the number of agents (implicit in the profile). 

We denote by $\vec a_{-i}$ the partial profile that includes all entries in $\vec a$ except $a_i$. 

A  $n$-agent \emph{facility location mechanism} in domain $\tup{\calX,d}$ (or simply \emph{a mechanism}) is a function $f:\calX^n\rightarrow \Delta(\calX)$, where $\Delta(\calX)$ is the set of distributions over $\calX$. We denote the resulting lottery of applying $f$ to profile $\vec a$ by $f_\vec a$. Mechanism $f$ is  \emph{deterministic} if $f_{\vec a}$ is  \rmr{change to $f_{\vec a}$} degenerated for any profile $\vec a$, in which case we denote $f_{\vec a}\in \calX$.
We denote the probability that mechanism $f$ selects $z$ on profile $\vec a$ by $f_{\vec a}(z)\in[0,1]$.  

When placing a facility on $z\in \calX$, an agent located at $a_i$ suffers a cost of $d(a_i,z)$. 
We denote by $c_i(\vec a,h)=E_{z\sim h}[d(a_i,z)]$ the expected cost of agent~$i$ in profile $\vec a$, when the facility is placed according to lottery $h$. 

The (utilitarian) \emph{social cost} of lottery $h$ in profile $\vec a$ is denoted by $SC(\vec a,h)= \sum_{i\leq n}c_i(\vec a,h)= E_{z\sim h}[\sum_{i\leq n}d(a_i,z)]$.

We omit the parameter $\vec a$ from the last two definitions when clear from context. We also abuse notation by writing $c_i(\vec a,z),SC(\vec a,z)$ for a specific location $z\in \calX$ rather than a lottery. 

We denote by $OPT(\vec a)=\inf_{z\in \calX}SC(\vec a,z)$ the optimal social cost (note that this is w.l.o.g. obtained in a deterministic location). 


\subsection{Common Mechanism Properties}
A mechanism $f$ is \emph{strategyproof} if for any profile $\vec a\in \calX^n$, any agent $i$, and any alternative report $a'_i\in \calX$, $c_i(\vec a,f_{\vec a}) \leq c_i(\vec a,f_{\vec a_{-i},a'_i})$ (i.e., $i$ does not gain in expectation).

A mechanism $f$ is \emph{ex-post strategyproof} if it is a lottery over strategyproof deterministic mechanisms. 
Note that ex-post strategyproofness implies strategyproofness, but not vice versa. 

%
%

A mechanism $f$ is \emph{peaks-only} if $f_{\vec a}(z)=0$ for all $z\notin \vec a$. That is, if the facility can only be realized on agents' locations. 

Mechanism $f$ \emph{dominates} mechanism $g$, if for any profile $\vec a$, $SC(\vec a,f_{\vec a})\leq SC(\vec a,g_{\vec a})$ and the inequality is strict for at least one profile. 

Finally, a mechanism $f$ has an approximation ratio of $\phi$, if for any profile $\vec a$, $SC(\vec a,f_{\vec a})\leq \phi \cdot OPT(\vec a)$. 
%
\paragraph{Familiar mechanisms}
The \emph{Random Dictator (RD) mechanism} selects each agent~$i$ with equal probability, and places the facility on $a_i$. Clearly RD is ex-post strategyproof, and it is also known to be group-strategyproof~\cite{alon2010walking} (that is, no subset of agents can gain by a joint deviation). Further, RD has an approximation ratio of $2-\frac2n$ (i.e., $\frac43$ for $n=3$ agents), and this is tight for any metric space with at least two distinct locations~\cite{alon2010strategyproof}. 

On one-dimensional spaces, where agent locations can be sorted, the deterministic \emph{median} mechanism simply picks the location of the median agent. The median mechanism is strategyproof and optimal~\cite{Moul80}. The median mechanism also extends to trees, maintaining both properties~\cite{SV04}.  

\section{Circles}
A circle is the simplest graph for which there is no median. We denote by $C_M$ the circle graph with $M$ equi-distant vertices $V$. Assume w.l.o.g. that agents are indexed in clockwise order. For a profile $\vec a\in V^n$, and two consequent agents $j,j+1$ (the addition is modulo $n$), we denote by $L_{\vec a}(a_j,a_{j+1})$ (or just $L(a_{j},a_{j+1})$ when the profile is clear from context) the length of the arc between these agents, that does not contain any other agent. When $L(a_j,a_{j+1})$ is not larger than a semicircle, then it also coincides with the distance $d(a_j,a_{j+1})$.  

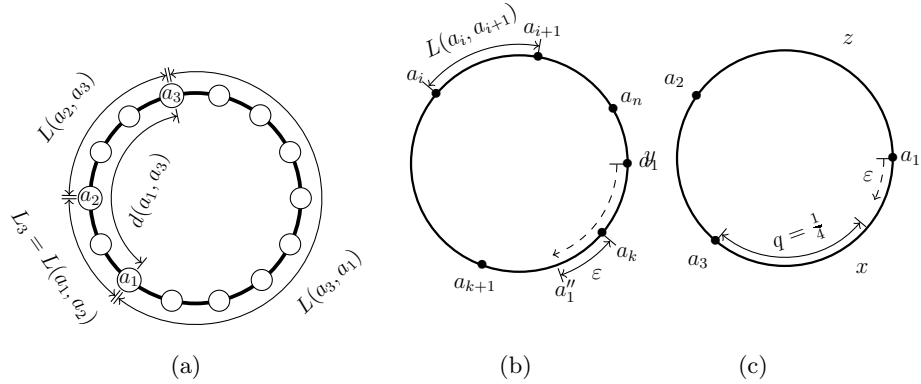
\begin{figure}[t]
\begin{subfigure}[b]{0.45\textwidth}

\tikzstyle{VertexStyle}=[circle,draw=black,fill=white,
inner sep=0pt,minimum size=8mm]
\centering

\tikzstyle{dot}=[circle,draw=black,fill=white,
inner sep=0pt,minimum size=4mm]
\tikzset{
  font={\fontsize{12pt}{12}\selectfont}}
\begin{tikzpicture}[scale=0.7,transform shape]

\draw[line width=0.5mm] (0,0) circle (2cm);

  \node at (-0:2cm) [dot] {};
\node at (-26:2cm) [dot] {};
\node at (-51:2cm) [dot] {};
\node at (-77:2cm) [dot] {};
\node at (-103:2cm) [dot] {};
\node at (-129:2cm) [dot] {$a_1$};
\node at (-154:2cm) [dot] {};
\node at (-180:2cm) [dot] {$a_2$};
\node at (-206:2cm) [dot] {};
\node at (-231:2cm) [dot] {};
\node at (-257:2cm) [dot] {$a_{3}$};
\node at (-283:2cm) [dot] {};
\node at (-308:2cm) [dot] {};
\node at (-334:2cm) [dot] {};

\draw[|<->|] (-129:2.4cm) arc (-129:-180:2.4cm);
\node[rotate=-55] at (-154:3cm) {$L_3=L(a_1,a_2)$};

\draw[|<->|] (-181:2.4cm) arc (-181:-257:2.4cm);
\node[rotate=55] at (-215:3cm) {$L(a_2,a_3)$};

\draw[|<->|] (-258:2.4cm) arc (-258:-488:2.4cm);
\node[rotate=45] at (-32:3cm) {$L(a_3,a_1)$};

\draw[|<->|] (-258:1.6cm) arc (-258:-129:1.6cm);
\node[rotate=70] at (-190:0.9cm) {$d(a_1,a_3)$};

\end{tikzpicture}
%

%
\caption{\label{fig:PCM}}
\end{subfigure}
\begin{subfigure}[b]{0.25\textwidth}

\tikzstyle{VertexStyle}=[circle,draw=black,fill=white,
inner sep=0pt,minimum size=8mm]
\centering

\tikzstyle{dot}=[circle,draw=black,fill=white,
inner sep=0pt,minimum size=4mm]
\tikzset{
  font={\fontsize{12pt}{12}\selectfont}}
\begin{tikzpicture}[scale=0.72,transform shape]

\draw[line width=0.3mm] (0,0) circle (2cm);

  \node at (-0:2.4cm)  {$a_1$};
	\node at (-0:2cm)  {$\bullet$};
	
\node at (-70:2.5cm) {$a''_1$};
\node at (-40:2.6cm)  {$a_k$};
	\node at (-40:2cm)  {$\bullet$};
\node at (-110:2.4cm)  {$a_{k+1}$};
	\node at (-110:2cm)  {$\bullet$};
\node at (-220:2.5cm)  {$a_i$};
	\node at (-220:2cm)  {$\bullet$};
\node at (-280:2.5cm)  {$a_{i+1}$};
	\node at (-280:2cm)  {$\bullet$};
\node at (30:2.4cm)  {$a_n$};
	\node at (30:2cm)  {$\bullet$};

\draw[|->,dashed] (0:1.8cm) arc (0:-70:1.8cm);

\draw[|<->|] (-40:2.2cm) arc (-40:-70:2.2cm);
\node at (-55:2.5cm) {$\varepsilon$};
\draw[|<->|] (-220:2.2cm) arc (-220:-280:2.2cm);
\node[rotate=25] at (-250:2.6cm) {$L(a_i,a_{i+1})$};

\end{tikzpicture}
%
%
\caption{\label{fig:PCD_SP}}
\end{subfigure}
\begin{subfigure}[b]{0.25\textwidth}
\tikzstyle{VertexStyle}=[circle,draw=black,fill=white,
inner sep=0pt,minimum size=8mm]
\centering
\tikzstyle{dot}=[circle,draw=black,fill=white,
inner sep=0pt,minimum size=4mm]
\tikzset{
  font={\fontsize{12pt}{12}\selectfont}}
\begin{tikzpicture}[scale=0.72,transform shape]
\def\rc{1.66}

\draw[line width=0.3mm] (0,0) circle (\rc*1.2cm);

  \node at (0:\rc*1.4cm)  {$a_1$};
	\node at (0:\rc*1.2cm) {$\bullet$};
\node at (-130:\rc*1.5cm) {$a_3$};
\node at (-130:\rc*1.2cm) {$\bullet$};
\node at (145:\rc*1.5cm)  {$a_2$};
\node at (145:\rc*1.2cm) {$\bullet$};

\node at (-90:\rc*1.6cm) {};
\node at (-55:\rc*1.5cm) {$x$};

\node at (62:\rc*1.5cm) {$z$};

\draw[|<->|] (-40:\rc*1.1cm) arc (-40:-130:\rc*1.1cm);
\node at (180:\rc*1.5cm) {$y$};
\node[rotate=20] at (-80:\rc*0.85cm) {$q=\frac14$};

\draw[dashed,|->] (0:\rc*1.1cm) arc (0:-25:\rc*1.1cm);
\node at (-12:\rc*0.95) {$\varepsilon$};

\end{tikzpicture}
%
%
\caption{\label{fig:QCD_SP}}
\end{subfigure}
\caption{Examples. (a) The circle $C_{14}$. Under PCD mechanism, the probabilities that the facility will be realized on $a_1, a_2$ and $a_3$, respectively, are $(\frac{3}{14},\frac{9}{14},\frac{2}{14})$. Under  PD, the probabilities are $(\frac{3}{10},\frac{5}{10},\frac{2}{10})$. Under $\frac14$-QCD, the probabilities are proportional to $((\frac{1}{4})^2 ,(\frac{9}{14})^2,(\frac{1}{4})^2)$, which gives us $(0.1161,0.7677,0.1161)$. The other two examples are used in the proof of Theorem~\ref{thm:PCD_SP} (b) and Case~I of Theorem~\ref{thm:QCD_SP} (c).  }
\end{figure}

We also define $L_i = L(a_{j},a_{j+1})$ where $j=i+\floor{n/2}$ (modulo $n$) to be the length of the arc that is ``facing'' agent~$i$ (although it may not be antipodal). For 3 agents this simply means that $L_1 = L(a_2,a_3), L_2=L(a_3,a_1)$, and $L_3 = L(a_1,a_2)$.  Also note that for 3 agents, the optimal location is always the agent facing the longest arc. See Fig.~\ref{fig:PCM}. 
 
\subsection{Proportional Distance}
\begin{definition}
The \emph{Proportional Circle Distance (PCD) mechanism} 
assigns the facility to each location $a_i$ w.p. $\frac{L_i}{\sum_{j\leq n}L_j}$. 
\end{definition}

\begin{theorem}\label{thm:PCD_SP}PCD is strategyproof for any odd $n$.
\end{theorem}
\begin{sproof}
Suppose that $a_1$ tries the manipulate by moving (w.l.o.g.) clockwise to $a'_1$. Note that the probability of selecting agent~$1$ is not affected. Thus the agent's gain comes from increasing the selection probability of a closer agent at the expense of a farther agent. On the other hand, the agent's cost increases proportionally to her distance from her true location, and we show that this factor is more prominent.   \QED
\end{sproof}

\rmr{
Note that another equivalent formulation of PCD mechanism is to remove a random edge from the circle and then run the median mechanism on the remaining line. This supposedly shows that PCD is ex-post strategyproof (since it randomizes over medians  and the median mechanism is strategyproof). However the ``removed'' edge still exists when agents consider their distance from the facility and thus an agent may have an ex-post manipulation.   }

For 3 agents, the PCD mechanism guarantees an approximation ratio of $\frac54=1.25$. This is not hard to show, but will also follow from stronger results in Section~\ref{sec:graphs}.  In Section~\ref{sec:more_than_3} we further discuss what we know when $n>3$.

\subsection{The Quadratic Distance Mechanism}
Since the optimal location with 3 agents is always the peak facing the longest arc, to improve the approximation ratio we must put more weight on peaks facing long arcs (at least in the ``bad'' instances). 

\begin{definition}
The \emph{$q$-Quadratic Circle Distance ($q$-QCD) mechanism} considers the arc lengths $L_1,L_2,L_3$.  It then assigns the facility to $a_i$ w.p. \emph{proportional to} $s_i=\max\{(L_i)^2,q^2\}$.
\end{definition}
That is, $q$ puts a lower bound on the probability that each agent is selected.
\begin{theorem}\label{thm:QCD_SP}
The $\frac{1}{4}$-QCD mechanism is strategyproof.
\end{theorem}
\begin{sproof}
	We denote  $x=L_2,z=L_3$ and $y=L_1$. We denote by $s_x,s_y,s_z$ the un-normalized weight assigned to the agent facing each respective arc, and by $p_i = \frac{s_i}{s}$ where $s=s_1+s_2+s_3$ the actual probability that $i$ is selected. Note that $p_x+p_y+p_z=1$. The notations are demonstrated on Fig.~\ref{fig:QCD_SP}
		
 The cost to agent~1 can be written as
	$$c_1 = p_x z + p_z x = \frac{s_x z + s_z x}{s_x+s_y+s_z}.$$
Consider a step of size $\eps$ by agent~1 towards agent~3. Intuitively, moving towards the far agent only increases its probability of selection and is thus never beneficial for agent~1. Thus w.l.o.g. $z\geq x \geq \eps$. 

The move changes the arc lengths from $(x,y,z)$ to $(x-\eps,y,z+\eps)$, and the cost changes accordingly to
\labeq{c1_move}
{c'_1 = p'_x z + p'_z x + p'_y \eps =  \frac{s_{x-\eps} z + s_{z+\eps} x +s_y \eps}{s_{x-\eps}+s_y+s_{z+\eps}}.}
Our general strategy is to write the new cost $c'_1$ as
\labeq{c'_1_gen}
{c'_1 = \frac{s_x z + s_z x  + \eps \gamma}{s_x+s_y+s_z + \eps \theta}=\frac{c_1 s  + \eps \gamma }{s + \eps \theta},}
where $\gamma,\theta\geq 0$. Then, we show that $\frac{\gamma}{\theta}\geq \frac{s_x z + s_z x}{s_x+s_y+s_z}(=c_1)$. This would conclude the proof, as it means that agent~1 does not gain:
\labeq{idea}
{c'_1 = \frac{s_{x-\eps} z + s_{z+\eps} x +s_y \eps}{s_{x-\eps}+s_y+s_{z+\eps}} \geq \frac{c_1 s  + \eps  c_1\theta}{s + \eps \theta} = \frac{c_1 (s+\eps\theta)}{s+\eps\theta} = c_1.}

 The exact values of $\gamma,\theta$ depend on whether $x-\eps\geq q$ (Case~I, see Fig.~\ref{fig:QCD_SP}), $x\geq q > x-\eps$ (Case~II), or $q>x$ (Case~III). We only show here Case~I, which captures most of the proof's ideas. The proofs of the other cases are similar, with some caveats. 
	
	Suppose first that $y\geq q=\frac14$ and that $z\leq \frac12$ (we later show this does not matter). Then $s_x = x^2, s_y = y^2, s_z = z^2$, and 
$$c_1 = p_x z + p_z x = \frac{x^2z+z^2x}{x^2+z^2+y^2}.$$
	After the move, we have $s'_x = (x-\eps)^2, s'_z =  (z+\eps)^2, s'_y = s_y= y^2$. Plugging into Eq.~\eqref{eq:c1_move},
	\begin{align*}
	c'_1 &=  \frac{(x-\eps)^2 z+(z+\eps)^2x + y^2 \eps}{(x-\eps)^2+(z+\eps)^2+y^2}\\
	& = \frac{x^2z -2\eps xz +\eps^2 z +z^2x +2\eps zx + \eps^2 x+ y^2 \eps}{x^2-2\eps x +\eps^2 + z^2+2\eps z +\eps^2+y^2}\\
	& = \frac{x^2z +z^2x + \eps(y^2+ \eps( z+ x))}{x^2+ z^2+y^2 +2\eps(z-x+\eps)} = \frac{c_1 s + \eps \gamma}{s + \eps\theta}.\\
	\end{align*}
	It is worthwhile to take a step back and consider what we got so far. Note that $\gamma$ in the nominator is always positive because the (linear) derivatives of the quadratic terms $s_x z, s_z x$ cancel out. This shows why using quadratic probabilities makes sense. However, this is not sufficient, since $\theta$ in the denominator is also positive, and when $s_y$ is too small (specifically, smaller than $\frac{1}{16}$) then the nominator grows too slowly to counter the increase in the denominator. This explains why we need the parameter $q$---to make sure that the manipulator is selected with sufficient probability to counter the benefit of the increased probability of the agent that is closer to $a_1$. 
	
	Going back to the technical proof, we need to show that
	$$\frac{\gamma}{\theta}=\frac{y^2+ \eps( z+ x)}{2(z-x+\eps)} \geq \frac{x^2z +z^2x}{x^2+ z^2+y^2}.$$
	Rearranging, we should prove that 
	
	\labeq{exp_1}
	{(y^2+ \eps( z+ x))(x^2+ z^2+y^2) - (x^2\underline{z} +z^2x)(2(z-x+\eps))}
	is non-negative. It is easy to see that this expression is monotonically increasing in $y$ (and $y\geq \frac14$ in this case). 
	It is a bit less easy to see (not shown here) that it is also monotonically increasing in $\eps$.
	%
	Thus it is sufficient to lower bound $(\frac{1}{16}+ x( z+ x))(x^2+ z^2+\frac{1}{16}) - (x^2z +z^2x) 2(z-x+x)$, or, equivalently,
	$$(\frac{1}{16}+ xz+ x^2)(x^2+ z^2+\frac{1}{16})-2z^2x(x+z).$$
	One can check that the minimum of this expression in the range $0\leq x\leq z\leq \frac12$ is exactly $0$ (at $z=\frac12,x=\frac14$).\footnote{We verified this with Wolfram Alpha.} Thus $\frac{\gamma}{\theta}\geq c_1$, and we are done by Eq.~\eqref{eq:idea}.
	
	Finally, suppose that $z>\frac12$. The only change is that the underlined $ z$ in Eq.~\eqref{eq:exp_1} would change to $x+y$ (which is smaller than $z$). This only increases the expression and would thus not make it negative. \QED
\end{sproof}

Since the inequality we get in Eq.~\eqref{eq:exp_1} is tight, the proof also shows that any $q$-QCD mechanism for $q<\frac14$ would not be strategyproof. 

\begin{proposition}\label{prop:QCD_approx}
The $\frac14$-QCD mechanism has an approximation ratio of $\frac76 \cong 1.166$, and this is tight.
\end{proposition}
\begin{proof}
Let $\vec a = (a_1,a_2,a_3)$ be a profile, and denote $x=d(a_1,a_2), y=d(a_2,a_3), z=d(a_1,a_3)$. We assume w.l.o.g.  $z\geq y\geq x$, thus the optimal point is $a_2$. The optimal social cost is $x+y$. 

We first argue that the approximation only becomes worse by moving $a_2$ to the mid point between  her neighbors. By decreasing $y$ to $y'=y-\eps$ and increasing $x$ to $x'=x+\eps$, $z$ remains the largest arc, so $a_2$ is still optimal and $x'+y'=x+y$ is still the optimal social cost. 
The social cost of the mechanism changes from $\frac{s_x (y+z) + s_y (x+z) + s_z(x+y)}{s_x+s_y+s_z}$ to $\frac{s'_x (y+z) + s'_y (x+z) + s_z(x+y)}{s'_x+s'_y+s_z}$. We have that $s'_x+s'_y\leq s_x+s_y$ since the new partition is more balanced. This means that the denominator weakly increases and the total weight $p_z$ given to the optimal point $a_2$ can only decrease. Among the two non optimal points, note that $a_3$ has the higher cost ($z+y\geq z+x$).  Now,  $s'_x \geq s_x$ so the relative weight of the worst point $a_3$ only increases. Thus the social cost weakly increases and the approximation ratio becomes worse. 

This means that we are left to find the worst instance among the instances with distances $(x,x,1-2x)$ for some $x\leq \frac13$. The optimum in such an instance is $2x$ whereas the social cost of $\frac14$-QCD is:
\begin{itemize}
	\item for $\frac13 \geq x\geq \frac14$, we have in particular that $1-2x\geq \frac13> \frac14$. Then 
	
	\vspace{-1mm}
	$$SC= \frac{2x^2(x+(1-2x))+(1-2x)^2 2x}{2x^2+(1-2x)^2} = \frac{2x-6x^2+6x^3 }{1-4x+6x^2}$$ 
	and the approximation ratio is $\frac{1-3x+3x^2  }{1-4x+6x^2}$. The derivative of this expression is negative for $x<\frac12$ so it is maximized at the bottom of the range, at $x=\frac14$. 
	\item for $x\leq\frac14$, we have that 
	
	\vspace{-4mm}
	$$SC = \frac{2(1/4)^2 3x+(1-2x)^2 2x}{2(1/4)^2+(1-2x)^2},$$
	and the approximation ratio is $\frac{3/16 +(1-2x)^2}{2/16+(1-2x)^2}$, which is increasing in $x$, so once again we obtain the maximum at $x=\frac14$.  
\end{itemize}
Plugging $x=\frac14$ to the expression of the approximation ratio above, we get that in the worst instance $\vec a=(0,\frac14,\frac12)$, $\frac14$-QCD obtains an approximation ratio of exactly $\frac{3/16 +(1/2)^2}{2/16+(1/2)^2}=\frac76$. \QED
\end{proof}

\subsection{Beyond 3 agents}\label{sec:more_than_3}
We already saw that the PCD mechanism is strategyproof for any odd $n$.
However, calculating its worst-case approximation ratio is more tricky. In particular,  the worst instance is \emph{not}  symmetric w.r.t. the optimal point (in contrast to 3 agents). In the limit, PCD and random dictator have the same approximation:

\begin{proposition}\label{prop:PCD_2}
When $n$ grows, the approximation ratio of PCD approaches $2$.
\end{proposition}
\begin{proof}
Let $n=2k+1$, and consider the profile in Fig.~\ref{fig:PCD_bad}, where $x=\frac{1}{4\sqrt k}$. The numbers inside the circle indicate the number of agents in each location.

 The optimal point is the bottom concentration, with a social cost of $c_1=k x + \frac12-x \leq \frac14\sqrt k+\frac12$. The social cost of the left point is $c_2 =k(\frac12-x) + k\frac12$, and of the right point is $c_3=kx+\frac12$. Thus
%
\begin{align*}
SC&(f^{PCD}(\vec a))=\frac12 c_1  + x c_2 + (\!\frac12-x\!) c_3
= -2k x^2 + (2k-1)x + \frac12 \\
&= -\frac18 + \frac12 \sqrt k - \frac1{4\sqrt k} + \frac12 \geq \frac12 \sqrt k, 
\end{align*}
and thus the approximation ratio is at least $\frac{\frac12 \sqrt k}{\frac14 \sqrt k+\frac12}> 2-\frac{8}{\sqrt n}$. \QED
\end{proof}

%

It is an open question  whether there is some mechanism (perhaps a variation of  $q$-QCD) that strictly beats 2 approximation for any $n$.   We believe that this is indeed the case but that would require simplifying the proof technique. 

\paragraph{Peak-only restrictions} We prove a weaker version of the following:
\begin{conjecture}\label{conj:peaks_only}
For any $n$, the best strategyproof mechanism is peaks-only.
\end{conjecture}
\begin{proposition}\label{prop:antipod}
For any $n$, the optimal strategyproof mechanism w.l.o.g. only places the facility either on peaks, or on points antipodal to peaks. 
\end{proposition}
\begin{sproof}
For a profile $\vec a=(a_1,\ldots,a_n)$, denote by $b_i$ the point antipodal to $a_i$, and let $A=\{a_1,\ldots,a_n,b_1,\ldots,b_n\}$. 
Suppose that in some some profile $\vec a$, the mechanism $f$ places the facility with some probability $p$ on point $\alpha\notin A$. Denote by $\beta,\gamma$ the nearest points to $\alpha$ from $A$ clockwise and counterclockwise, respectively. Let $x=d(\alpha,\beta), y=d(\alpha,\gamma)$ (see Fig.~\ref{fig:antipod}). 

We define a mechanism $f'$ that is identical to $f$, except that it ``splits'' the probability mass $p$ of $\alpha$ between the adjacent points $\beta,\gamma$:  it sets ${f'}_{\vec a}(\alpha)=0$; $f'_{\vec a}(\beta) = f_{\vec a}(\beta) + p\frac{y}{x+y}$; and $f'_{\vec a}(\gamma) = f_{\vec a}(\gamma) + p\frac{x}{x+y}$.

We claim that for any agent~$i$, $c_i(\vec a,f_{\vec a}) = c_i(\vec a, f'(\vec a))$. This would show both that $f'$ is strategyproof (since $f$ is) and that $SC(\vec a,f_{\vec a})=SC(\vec a, f'(\vec a))$ for all $\vec a$. 

Indeed, consider some agent placed at $a_i$. From the three points $\alpha,\beta,\gamma$, the one farthest from $a_i$ cannot be $\alpha$, since this would mean that $b_i$ (the point antipodal to $a_i$) is strictly in the open interval $(\beta,\gamma)$, whereas by construction there are no more points from $A$ in this interval. 
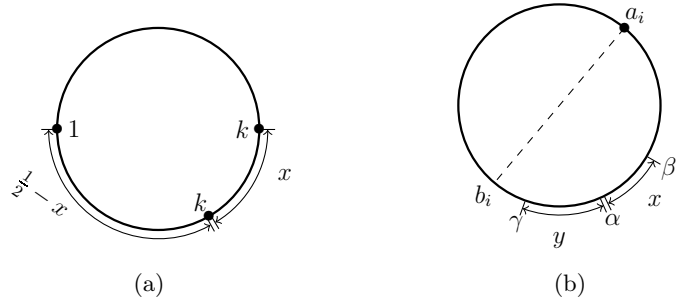
\begin{figure}[t]
\centering
\begin{subfigure}[b]{0.45\textwidth}

\tikzstyle{VertexStyle}=[circle,draw=black,fill=white,
inner sep=0pt,minimum size=8mm]
\centering

\tikzstyle{dot}=[circle,draw=black,fill=white,
inner sep=0pt,minimum size=4mm]
\tikzset{
  font={\fontsize{12pt}{12}\selectfont}}
\begin{tikzpicture}[scale=0.8,transform shape]
\def\rcc{1.4}
\draw[line width=0.3mm] (0,0) circle (\rcc*1.2cm);

  \node at (-0:\rcc*1cm)  {$k$};
	  \node at (-0:\rcc*1.2cm)  {$\bullet$};
\node at (-60:\rcc*1cm) {$k$};
\node at (-60:\rcc*1.2cm)  {$\bullet$};
\node at (-180:\rcc*1cm)  {$1$};
\node at (-180:\rcc*1.2cm)  {$\bullet$};

\draw[|<->|] (0:\rcc*1.3cm) arc (-0:-59:\rcc*1.3cm);
\node at (-20:\rcc*1.6cm) {$x$};

\draw[|<->|] (-61:\rcc*1.3cm) arc (-61:-180:\rcc*1.3cm);
\node[rotate=-35] at (-150:\rcc*1.6cm) {$\frac12-x$};

\end{tikzpicture}
%

%
\caption{\label{fig:PCD_bad}}
\end{subfigure}
\begin{subfigure}[b]{0.45\textwidth}

\tikzstyle{VertexStyle}=[circle,draw=black,fill=white,
inner sep=0pt,minimum size=8mm]
\centering

\tikzstyle{dot}=[circle,draw=black,fill=white,
inner sep=0pt,minimum size=4mm]
\tikzset{
  font={\fontsize{12pt}{12}\selectfont}}
\begin{tikzpicture}[scale=0.8,transform shape]
\def\rc{1.4}

\draw[line width=0.3mm] (0,0) circle (\rc*1.2cm);

  \node at (50:\rc*1.4cm)  {$a_i$};
	\node at (50:\rc*1.2cm)  {$\bullet$};
\node at (-30:\rc*1.5cm) {$\beta$};
\node at (-65:\rc*1.5cm)  {$\alpha$};
\node at (-110:\rc*1.5cm)  {$\gamma$};
\node at (-130:\rc*1.4cm)  {$b_i$};

\draw[|<->|] (-30:\rc*1.3cm) arc (-30:-64:\rc*1.3cm);
\node at (-45:\rc*1.6cm) {$x$};

\draw[|<->|] (-66:\rc*1.3cm) arc (-66:-110:\rc*1.3cm);
\node at (-90:\rc*1.6cm) {$y$};

\draw[dashed] (50:\rc*1.2cm) -- (-130:\rc*1.2cm) ;

\end{tikzpicture}
%

%
\caption{\label{fig:antipod}}
\end{subfigure}
\caption{\label{fig:examples_2} Figures used in the proofs of Prop.~\ref{prop:PCD_2} (a) and Prop.~\ref{prop:antipod} (b).}
\end{figure}
Thus w.l.o.g. $d(a_i,\beta)<d(a_i,\alpha)<d(a_i,\gamma)$ (see figure). We omit the rest of the proof, which is not hard.\QED
\end{sproof}

\section{Beyond Circles}\label{sec:graphs}
\begin{definition}
The \emph{Proportional  Distance (PD) mechanism} for three agents selects each $a_i$ ($i\in \{1,2,3\}$) with probability proportional to the distance between the other pair of agents.
\end{definition}

 Note that for three agents on a circle, PD and PCD coincide when the agents are not all on the same semicircle, and otherwise PCD gives higher probability to the ``middle'' agent (which is optimal). Therefore PCD dominates PD. See Fig.~\ref{fig:PCM} for an example.  It is also not hard to show that PD dominates RD on any metric space. 
In particular, this means that $SC(f^{PD}_{\vec a})\leq \frac43 OPT(\vec a)$ on any graph.

\begin{theorem}\label{thm:PD_SP}
The PD mechanism is strategyproof in expectation for 3 agents in any metric space (in particular on any graph).
\end{theorem}
In contrast to Theorem~\ref{thm:PCD_SP}, the proof is rather technical and is thus omitted.

\begin{observation}\label{ob:star}
The approximation ratio of any peaks-only mechanism (regardless of its incentive properties) on a general graph is at least $\frac43$ ($2-\frac2n$ for general $n$). 
\end{observation}
To see why, consider a star graph with $n$ leafs, each containing one agent. 

\begin{proposition}\label{prop:from_peak}
Let $f$ be any  peaks-only mechanism. Then for any profile $\vec a\in V^3$, we have that $SC(f^{PD}_{\vec a})\leq \frac54 SC(f_{\vec a})$, and this bound is tight. 
\end{proposition}
\begin{proof}
Consider the distances between pairs $x\leq y \leq z$. W.l.o.g. we can denote $x+y=1$.
By triangle inequality, $z\leq x+y=1$. 
 The optimal peak location yields a cost of $x+y=1$. The PD mechanism yields a cost of 
\begin{align*}
SC(f^{PD}) &= \frac{x(y+z)}{x+y+z} + \frac{y(x+z)}{x+y+z} + \frac{z(x+y)}{x+y+z} = \frac{2xy+xz + yz + z}{1+z}\\
& =2\frac{xy+ z}{1+z}\leq 2\frac{xy+ 1}{1+1} = xy+1 \leq (0.5)^2+1 = \frac54,
\end{align*}
as required.

For tightness, consider any domain that contains three points $a_1,a_2,a_3$ such that $a_2$ is in the middle between $a_1$ and $a_3$ (e.g., a line). If there is one agent on each point then $x=d(a_1,a_2)=d(a_2,a_3)= y=0.5$ whereas $z=d(a_1,a_3)=x+y=1$. Then $SC(f^{PD}(\vec a))=2\frac{xy+ z}{1+z}=1.25=1.25 OPT(\vec a)$, as the optimal peaks-only mechanism will select $a_2$. \QED
\end{proof}

Since the optimal point on a circle is always a peak, and since PCD dominates PD, we get the following. 
\begin{corollary}\label{cor:PD_circle}
 For $3$ agents on a circle, the PD and PCD mechanisms have an approximation ratio of $\frac54$, and this is tight.
\end{corollary}

\begin{remark} Since $d(a_1,a_2) + d(a_2,a_3) + d(a_3,a_1)$ is a constant $D$ given the profile, the PD mechanism selects each agent $i$ with probability proportional to $D-SC(a_i)$. This allows us to easily extends the PD mechanism to any $n$, and it remains an open question whether the PD mechanism remains strategyproof. Recall however that already on the circle, PCD dominates PD, and is not asymptotically better than random dictator.

In \cite{escoffier2011strategy}, the authors suggest a randomized mechanism for placing $n-1$ facilities based on a similar idea: they place facilities on all agents except one (assuming all locations are distinct), where the placement omitted location $a_i$ is selected with probability \emph{inversely proportional} to the social cost of this placement (which in their case is just the distance to the closest agent $j\neq i$), and show it is strategyproof for any $n$ and any metric space. Further, the inversely proportional mechanism is asymptotically better than randomly omitting a facility, which cannot guaranty any bounded approximation.  Another mechanism that uses a similar proportional lottery (for two facilities) is in \cite{lu2010asymptotically}.

This suggest another possible direction for a the single facility problem (or perhaps to more general problems), by considering various probabilities that are proportional to some decreasing function of the social cost.\footnote{Note that using the reciprocal of the social cost (as in \cite{escoffier2011strategy}) would lead to a poor outcome in the single facility problem.}
\end{remark}

\subsection{Lower Bounds via Linear Programming}\label{sec:lower}
An immediate corollary from Prop.~\ref{prop:antipod}, is that in that any upper bound on continuous circles implies the same upper bound on any finite circle with an even number of nodes, and thus any lower bound on a finite circle of any even size (or any size if Conj.~\ref{conj:peaks_only} is true) implies a lower bound for continuous circles. 

It is well known that mechanism design problems for finite domains can be written as linear programs~\cite{conitzer2002complexity}. Automated mechanism design had also been applied to facility location problems, for one or more facilities on a line~\cite{narasimhan2016automated,golowich2018deep}. Due to the specifics of the problems they considered, they used advanced machine learning techniques rather than linear programming. 

For a given graph $(V,E)$, finding the optimal randomized strategyproof mechanism for three agents can be written as a simple linear optimization program as follows. 
There are $|V|^4+1$ variables:  $(p_{\vec a,z})_{\vec a\in V^3,z\in V}$, where $p_{\vec a,z}=f_{\vec a}(z)$ is the probability that the facility is placed on $z$ in profile $\vec a$; and $\alpha\in \mathbb R$ which is the approximation factor. 
The optimization goal is simply to minimize $\alpha$. 
There are four types of constraints:
\begin{enumerate}
	\item Feasibility constraints: $p_{\vec a,z}\geq 0$ for all $\vec a\in V^3,z\in V$;
	\item Probability constraints: $\sum_{z\in V}p_{\vec a,z} = 1$ for all $\vec a\in V^3$;
	\item Incentive constraints: For every profile $\vec a\in V^3$, any agent $i\in \{1,2,3\}$, and any alternative location $a'_i\in V$, we want to enforce the constraint $c_i(\vec a,f_{\vec a})\leq c_i(\vec a,f_{(\vec a_{-i},a'_i)})$. This can be written as the following linear inequality over $2|V|$ variables:
	$$\sum_{z\in V} d(z,a_i) p_{\vec a,z} \leq \sum_{z\in V} d(z,a_i) p_{(\vec a_{-i},a'_i),z}.$$
	\item Approximation constraints: For every profile $\vec a\in V^3$, 
	we want to enforce the approximation $SC(\vec a,f_{\vec a})\leq \alpha \cdot OPT(\vec a)$. Since $OPT(\vec a)=\min_{z\in V}\sum_{i\in \{1,2,3\}}d(z,a_i)$ can be computed once for each profile, the approximation constraint can also be written as a linear inequality:
	$$\sum_{i\in \{1,2,3\}}\sum_{z\in V}d(z,a_i) p_{\vec a,z} \leq \alpha \cdot \min_{z\in V}\sum_{i\in \{1,2,3\}}d(z,a_i).$$
\end{enumerate}
In total, we get a bit more than $3|V|^4$ linear constraints. This is feasible for small graphs with commercial solvers, especially such that handle well sparse constraint matrices  (we used Matlab's \url{linprog} function). 

\rmr{One may also decide to drop some of the constraints, in which case the program will still return a correct lower bound that may not correspond to a valid mechanism. }

\begin{figure}[t]

\tikzstyle{VertexStyle}=[circle,draw=black,fill=white,
inner sep=0pt,minimum size=8mm]
\centering
\tikzset{
  font={\fontsize{12pt}{12}\selectfont}}
\begin{tikzpicture}[scale=0.7,transform shape]
  \Vertex[x=0,y=0,L=$v_1$]{v1}
	\Vertex[x=3,y=0,L=$v_2$]{v2}
	\Vertex[x=6,y=0,L=$v_3$]{v3}
	\Vertex[x=0,y=3,L=$v_4$]{v4}
	\Vertex[x=3,y=3,L=$v_5$]{v5}
	\Vertex[x=6,y=3,L=$v_6$]{v6}
	
  
  \tikzstyle{EdgeStyle}=[thick]
 	\Edge(v1)(v4)
	\Edge(v2)(v5)
	\Edge(v3)(v6)
	 \tikzstyle{EdgeStyle}=[dashed]
	\Edge(v1)(v5)
	\Edge(v1)(v6)
	\Edge(v2)(v4)
	\Edge(v2)(v6)
\Edge(v3)(v4)
\Edge(v3)(v5)
%
%

\end{tikzpicture}
%

\caption{\label{fig:graph_LB}A graph for which the best approximation ratio is $\frac{13}{12}$. The three solid edges have length~1, all dashed edges have length~2. }
\end{figure}
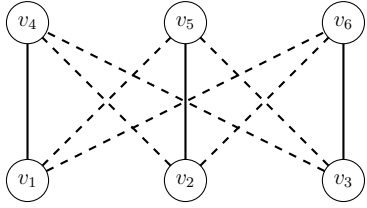

\begin{theorem}\label{thm:LB_graph}
There is no strategyproof mechanism for arbitrary graphs whose approximation ratio  is better than $\frac{13}{12}\cong 1.0833$.
\end{theorem}
\begin{proof}By coding the graph in Fig.~\ref{fig:graph_LB}, and using Matlab to solve the corresponding linear program. \QED
\end{proof}

\paragraph{Small circles}
\begin{lemma}\label{lemma:sym}For any strategyproof [peaks-only] mechanism $f$ on the circle, there is a neutral and anonymous strategyproof [peaks-only]  $g$,\footnote{A mechanism is \emph{neutral} if it is invariant to renaming of vertices, and \emph{anonymous} if it is invariant to renaming of agents.} such that $\max_{\vec a}SC(g_{\vec a}) \leq \max_{\vec a'}SC(f_{\vec a'})$. 
\end{lemma}
\begin{proof}
Mechanism $g$ simply selects a permutation over agents uniformly at random, and direction+rotation for the circle uniformly at random, thereby mapping profile $\vec a$ to $\hatv a$. Then, it runs $f$ on $\hatv a$ and maps back the outcome. Since this is a lottery over strategyproof mechanisms, it must also be strategyproof. It is also easy to see that if $f$ is peaks-only then so is $g$. Finally, for any profile $\vec a$, $SC(g_{\vec a})$ is averaging over several variations of $SC(f_{\hatv a})$, all of which are bounded by    $\max_{\vec a'}SC(f_{\vec a'})$. \QED
\end{proof}
\begin{figure}
\begin{center}
\includegraphics[scale=0.45]{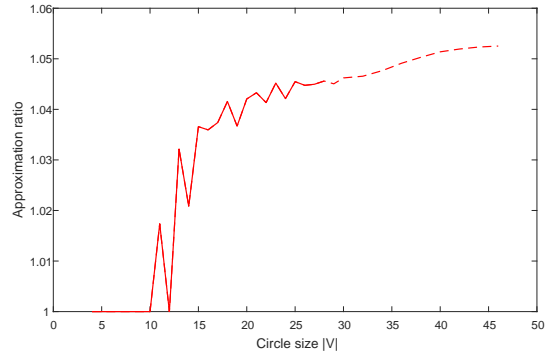}
\caption{\label{fig:LB_circle} The worst-case approximation ratio of the optimal 3-agent facility location mechanism on a circle with up to 44 vertices. The dashed part is computed only for peaks-only mechanisms on even $M$.}
\end{center}
\end{figure}

\begin{theorem}\label{thm:LB_circle}
There is no strategyproof mechanism for circle graphs whose approximation ratio  is better than $1.0456$. If we add the peaks-only requirement, the lower bound is $1.0523$.
\end{theorem}
To prove the theorem, we coded two linear programs: one that computes the optimal mechanism, and one that computes the optimal peaks-only mechanism.  Since the number of variables for a circle with $M$ vertices is $M^4$  (or $3M^3$ for peaks-only mechanisms) increases too fast for efficiently solving except for very small graphs, we applied the following improvements:
\begin{itemize}
	\item By Lemma~\ref{lemma:sym}, it is sufficient to check mechanisms that are neutral. We thus fixed the location of the first agent, which reduces the number of variables by a factor of $M$. 
	\item Also by Lemma~\ref{lemma:sym}, it is sufficient to check mechanisms that are anonymous. This allows us to add many symmetry constraints (both within profiles and between profiles) that effectively reduce the number of variables even more.
	\item By Prop.~\ref{prop:antipod}, it is sufficient to consider mechanisms that place the facility on one of the 6 peaks or anti-peaks.
\end{itemize}
This enables us to solve the obtained program for  all mechanism on circles up to $M=28$, and the program for peaks-only mechanisms for circles up to $M=44$. We note that the worst-case approximation bounds in both programs are the same for any $|V|\leq 28$, which supports Conjecture~\ref{conj:peaks_only}, but leaves the proof as a challenge.
 The worst-case approximation ratios of the optimal mechanism for finite circles are shown in Figure~\ref{fig:LB_circle}. It is non-monotone  due to parity effects. 

It remains an open question whether there is a better mechanism than the $\frac14$-QCD mechanism for circles of arbitrary size, and what is the best approximation ratio that can be guaranteed. While we improved the upper bound from $\frac43$ to $\frac76$, and the lower bound from $1$ to the bounds in Theorem~\ref{thm:LB_circle}, there is still a non-negligible gap. 

\section{Discussion}
Table~\ref{tab:results_r} summarizes our results for randomized mechanisms, and put them in the context of known bounds. 
It remains an open question whether the upper bound of $\frac43$ ($2-\frac2n$ for general $n$) is tight, and in particular whether general graphs are more difficult than circles. 

The effect of the circle size on the available strategyproof mechanisms was  evident in \cite{DFMN:2012:EC}. There, they showed (also using a computer search) a sharp dichotomy, where up to a certain size there are deterministic anonymous mechanisms, and above that size any strategyproof onto mechanism must be near-dictatorial. With randomized mechanisms, we see a more gradual effect.

The mechanisms we present seem quite specific to the problem at hand. Thus a natural question is what can be the takeaway messages for readers from the broader community of algorithmic game theory? We believe there are two. 

First, the idea of focusing on the \emph{derivative} of assignment probabilities as agents change their reported values. In the case of facility location,  misreporting a value (say, by $\eps$) causes the manipulator direct harm that is linear in $\eps$, but may change the outcome probabilities in a way that still makes the manipulation beneficial.  However, since the benefit is proportional to the \emph{change} in probabilities (i.e., to their derivatives), using  quadratic probabilities (whose derivatives are linear) puts the harm and benefit on the same scale. It is then left to the designer to tweak the parameters of the mechanism so as to make sure that the gain of a manipulator never exceeds the harm. Therefore, while the $q$-QCD mechanism seems more complicated than PCD and is more difficult to technically analyze, in a sense it is the result of a more structured and general approach to the problem, whereas PCD is a nice curiosity that happens to work.

The second idea is the combination of analytic and computational tools for solving a difficult design problem. While in some cases (e.g. in the analysis of our PD and PCD mechanisms) all the terms in the equations nicely cancel out to leave us with a clean proof, this is not always so. On the other hand, fully automated mechanism design~\cite{conitzer2002complexity} typically explodes with the size of the problem and leaves us with a solution that cannot be easily explained, modified or adapted to similar problems. This is true even for our linear programming approach in Section~\ref{sec:lower}. However, one can come up with a specific or parametrized class of mechanisms, and use the computer capabilities to prove certain difficult inequalities, optimize parameters, or test various conjectures before setting out to prove them analytically. A similar combined approach  has been applied e.g. in auctions~\cite{guo2010computationally}, albeit with very different mechanisms.
 


\begin{table*}[t]
\begin{center}
\begin{tabular}{|l||c|c|}
\hline
  metric space  &  Any  & Circle\\
	\hline
	Random Dictator   & $\frac43\cong 1.333$ (*)  &  $\frac43\cong 1.333$  (*) \\
	Proportional [Circle] Distance & $\frac43\cong 1.333$ (\#) & $\frac54 = 1.25$ (Cor.~\ref{cor:PD_circle}) \\
	$\frac14$-Quadratic Circle Distance & -     & $\frac76 \cong 1.166$ (Thm.~\ref{thm:QCD_SP},Prop.~\ref{prop:QCD_approx})\\
	\hline
best UB &  $1.333$ (RD/PD) & $1.166$ ($\frac14$-QCD)  \\
LB (peaks-only) &  $1.333$ (Ob.~\ref{ob:star})   &  $1.0523$ (Thm.~\ref{thm:LB_circle}) \\
LB &   $\frac{13}{12}\cong 1.0833$ (Prop.~\ref{thm:LB_graph})   & $1.0456$ (Thm.~\ref{thm:LB_circle})\\
\hline
\end{tabular}
\caption{\label{tab:results_r}A summary of approximation bounds for 3-agent randomized mechanisms. (*) - from \cite{alon2010strategyproof}. (\#) - obtains $\frac54=1.25$ approximation from best peak (Prop.~\ref{prop:from_peak}).}
\end{center}
\end{table*}

We leave many open questions for future research. In particular, whether the PD and QCD mechanisms can be generalized for more agents,  and whether there are classes of graphs that are inherently more difficult than circles.   
\clearpage
\bibliographystyle{abbrvnat}
\bibliography{abbshort,ultimate} 
\clearpage
\onecolumn
\appendix
\section{PCD}
\begin{rtheorem}{thm:PCD_SP}PCD is strategyproof for any odd $n$.
\end{rtheorem}
\begin{proof} 
Suppose that $a_1$ tries the manipulate by moving (w.l.o.g.) clockwise to $a''_1$. Let $k$ be the first index such that $a''_1>a_k$ and $a''_1$ is a beneficial manipulation (it is possible that $k=1$). Denote $a'_1=a_k$ and $\vec a'=(a'_1,a_2,\ldots,a_n), \vec a''=(a''_1,a_2,\ldots,a_n)$.  Denote $h=f_{\vec a}, h'=f_{\vec a'}$ and $h''=f_{\vec a''}$, where $f$ is the PCD mechanism.   Next, denote $\eps = a''_1-a'_1$, and consider the step where agent~1 moves from $a'_1$ to $a''_1$ (see Fig.~\ref{fig:ex}, left). 
This move changes the outcome from $h'$ to $h''$ and has two effects. 
First, it affects the selection probabilities (only) of $a_{i},a_{i+1}$ where $i=(n-1)/2+k-1$, and w.l.o.g. $a_i$ is closer to $a_1$ (otherwise the move is not beneficial). Second, it inflicts a \underline{cost} of $\eps$ w.p. $L(a_i,a_{i+1})$ (that is, in all realizations where agent~1 is selected). We need to show that the expected gain when moving from $a'_1$ to $a''_1$ is upper bounded by the expected cost.

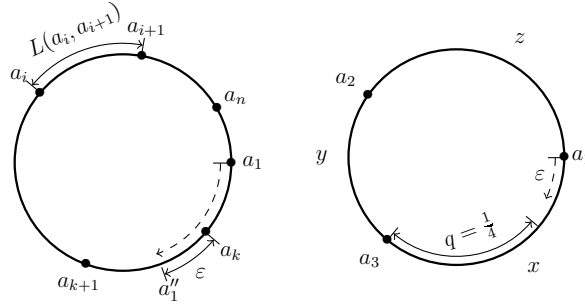
\begin{figure}[t]

\tikzstyle{VertexStyle}=[circle,draw=black,fill=white,
inner sep=0pt,minimum size=8mm]
\centering

\tikzstyle{dot}=[circle,draw=black,fill=white,
inner sep=0pt,minimum size=4mm]
\tikzset{
  font={\fontsize{12pt}{12}\selectfont}}
\begin{tikzpicture}[scale=0.72,transform shape]

\draw[line width=0.3mm] (0,0) circle (2cm);

  \node at (-0:2.4cm)  {$a_1$};
	\node at (-0:2cm)  {$\bullet$};
	
\node at (-70:2.5cm) {$a''_1$};
\node at (-40:2.6cm)  {$a_k$};
	\node at (-40:2cm)  {$\bullet$};
\node at (-110:2.4cm)  {$a_{k+1}$};
	\node at (-110:2cm)  {$\bullet$};
\node at (-220:2.5cm)  {$a_i$};
	\node at (-220:2cm)  {$\bullet$};
\node at (-280:2.5cm)  {$a_{i+1}$};
	\node at (-280:2cm)  {$\bullet$};
\node at (30:2.4cm)  {$a_n$};
	\node at (30:2cm)  {$\bullet$};

\draw[|->,dashed] (0:1.8cm) arc (0:-70:1.8cm);

\draw[|<->|] (-40:2.2cm) arc (-40:-70:2.2cm);
\node at (-55:2.5cm) {$\varepsilon$};
\draw[|<->|] (-220:2.2cm) arc (-220:-280:2.2cm);
\node[rotate=25] at (-250:2.6cm) {$L(a_i,a_{i+1})$};

\end{tikzpicture}
%
%
\tikzstyle{VertexStyle}=[circle,draw=black,fill=white,
inner sep=0pt,minimum size=8mm]
\centering
\tikzstyle{dot}=[circle,draw=black,fill=white,
inner sep=0pt,minimum size=4mm]
\tikzset{
  font={\fontsize{12pt}{12}\selectfont}}
\begin{tikzpicture}[scale=0.72,transform shape]
\def\rc{1.66}

\draw[line width=0.3mm] (0,0) circle (\rc*1.2cm);

  \node at (0:\rc*1.4cm)  {$a_1$};
	\node at (0:\rc*1.2cm) {$\bullet$};
\node at (-130:\rc*1.5cm) {$a_3$};
\node at (-130:\rc*1.2cm) {$\bullet$};
\node at (145:\rc*1.5cm)  {$a_2$};
\node at (145:\rc*1.2cm) {$\bullet$};

\node at (-90:\rc*1.6cm) {};
\node at (-55:\rc*1.5cm) {$x$};

\node at (62:\rc*1.5cm) {$z$};

\draw[|<->|] (-40:\rc*1.1cm) arc (-40:-130:\rc*1.1cm);
\node at (180:\rc*1.5cm) {$y$};
\node[rotate=20] at (-80:\rc*0.85cm) {$q=\frac14$};

\draw[dashed,|->] (0:\rc*1.1cm) arc (0:-25:\rc*1.1cm);
\node at (-12:\rc*0.95) {$\varepsilon$};

\end{tikzpicture}
%
%
\caption{\label{fig:ex}Examples used in proof of Theorem~\ref{thm:PCD_SP} (left) and Case~I of Theorem~\ref{thm:QCD_SP} (right).}
\end{figure}

{\small
\begin{align*}
&gain =  (h'(\! a_i\!)\!-\! h''(\! a_i\! ))d(\! a_i,a_1\!) +  (h'(\! a_{i+1}\!)\!-\! h''(\! a_{i+1}\!))d(\! a_{i+1},a_1\! )\\
&=  \eps d(a_i,a_1) + (-\eps)d(a_{i+1},a_1) = \eps (d(a_i,a_1)-d(a_{i+1},a_1))\\
&\leq \eps d(a_i,a_{i+1}) \leq \eps L(a_i,a_{i+1})= cost \tag{triangle inequality} \\
\end{align*}
}

Thus, $c_1(\vec a,h'')= c_1(\vec a,h')+cost-gain \geq c_1(\vec a,h') \geq c_1(\vec a,h),$
 where the last inequality is by our  minimality assumption.  
\end{proof}

\section{$q$-QCD}
\begin{rtheorem}{thm:QCD_SP}
The $\frac{1}{4}$-QCD mechanism is strategyproof.
\end{rtheorem}
\begin{proof}
	We denote by $x,z$ the lengths of the arcs adjacent to $a_1$, and by $y$ the arc facing $a_1$. We denote by $s_x,s_y,s_z$ the un-normalized weight assigned to the agent facing each respective arc, and by $p_i = \frac{s_i}{s_1+s_2+s_3}$ the actual probability that $i$ is selected. Note that $p_x+p_y+p_z=1$. 
		
		The cost to agent~1 can be written as
	$$c_1 = p_x z + p_z x = \frac{s_x z + s_z x}{s_x+s_y+s_z}.$$
Consider a step of size $\eps\leq x$ by agent~1. This changes the arc lengths from $(x,y,z)$ to $(x-\eps,y,z+\eps)$.
The cost changes accordingly to
$$c'_1 = p'_x z + p'_z x + p'_y \eps =  \frac{s_{x-\eps} z + s_{z+\eps} x +s_y \eps}{s_{x-\eps}+s_y+s_{z+\eps}}.$$
 Note that $s_{x-\eps}\leq s_x, s_{z+\eps}\geq s_z$. Also, the denominator is larger as the partition of the circle is more unbalanced.
Moving towards the farther agent always increases the cost (as it increases the nominator, and decreases the denominator of $c'_1$), so it cannot be a manipulation.

For the same reason, if $\eps>x$ then moving from $a_1+x$ to $a_1+\eps$ only hurts agent~1, since she is moving towards the far agent.

Thus w.l.o.g. $z\geq x \geq \eps$. Our general strategy is to write the new cost $c'_1$ as
\labeq{c'_1_gen_a}
{c'_1 = \frac{s_x z + s_z x  + \eps \gamma}{s_x+s_y+s_z + \eps \theta}=\frac{c_1 s  + \eps \gamma }{s + \eps \theta},}
where $\gamma,t\geq 0$. Then, we show that $\frac{\gamma}{\theta}\geq \frac{s_x z + s_z x}{s_x+s_y+s_z}(=c_1)$. This means that 
$$c'_1 \geq \frac{c_1 s  + \eps  c_1\theta}{s + \eps \theta} = \frac{c_1 (s+\eps\theta)}{s+\eps\theta} = c_1.$$

 The change in probabilities depends on the following cases. 
\begin{description}
	\item[Case~1:] $x-\eps\geq q$. Suppose first that $y\geq q=\frac14$ and that $z\leq \frac12$ (we later show this does not matter). Then $p_x \sim x^2, p_y \sim y^2, p_z \sim z^2$; and $p'_x\sim (x-\eps)^2, p'_z\sim (z+\eps)^2, p'_y \sim y^2$.
$$c_1 = p_x z + p_z x = \frac{x^2z+z^2x}{x^2+z^2+y^2}.$$
	After the move, we have 
	\begin{align*}
	c'_1 &= p'_x z+ p'_z x + p'_y \eps =  \frac{(x-\eps)^2 z+(z+\eps)^2x + y^2 \eps}{(x-\eps)^2+(z+\eps)^2+y^2}\\
	& = \frac{x^2z -2\eps xz +\eps^2 z +z^2x +2\eps zx + \eps^2 x+ y^2 \eps}{x^2-2\eps x +\eps^2 + z^2+2\eps z +\eps^2+y^2}\\
	& = \frac{x^2z +z^2x + \eps(y^2+ \eps( z+ x))}{x^2+ z^2+y^2 +2\eps(z-x+\eps)}.\\
	\end{align*}
	So in this case we need to show that
	$$\frac{\gamma}{\theta}=\frac{y^2+ \eps( z+ x)}{2(z-x+\eps)} \geq \frac{x^2z +z^2x}{x^2+ z^2+y^2}.$$
	Rearranging, we should prove that 
	
	\labeq{exp_1_a}
	{(y^2+ \eps( z+ x))(x^2+ z^2+y^2) - (x^2\boldsymbol{z} +z^2x)(2(z-x+\eps))}
	is non-negative. It is easy to see that this expression is monotonically increasing in $y$ (and $y\geq \frac14$ in this case). 
	It is a bit less easy to see that it is also monotonically increasing in $\eps$:
	\begin{align*}
	&(y^2+ \eps( z+ x))(x^2+ z^2+y^2) - (x^2z +z^2x)(2(z-x+\eps)) \\
	&=y^2(x^2+ z^2+y^2) -(x^2z +z^2x)2(z-x) + \eps[( z+ x)(x^2+ z^2+y^2) - 2(x^2z +z^2x)]\\
	&=y^2(x^2+ z^2+y^2) -(x^2z +z^2x)2(z-x) + \eps[zx^2+x^3+xz^2+z^3+ ( z+ x)y^2) - 2(x^2z +z^2x)]\\
	&\geq y^2(x^2+ z^2+y^2) -(x^2z +z^2x)2(z-x) + \eps[(x^2x+z^2z) - (x^2z +z^2x)],\\
	\end{align*}
 since the term in square brackets is nonnegative due to $z\geq x,z^2\geq x^2$. 
	Thus it is sufficient to lower bound
	$$(\frac{1}{16}+ x( z+ x))(x^2+ z^2+\frac{1}{16}) - (x^2z +z^2x) 2(z-x+x) = (\frac{1}{16}+ xz+ x^2))(x^2+ z^2+\frac{1}{16})-2z^2x(x+z).$$
	One can check that the minimum of this expression in the range $0\leq x\leq z\leq \frac12$ is exactly $0$ (at $z=\frac12,x=\frac14$).\footnote{We verified this with Wolfram Alpha.} Thus
	\begin{equation}\label{eq:c_1}
		c'_1 \geq \frac{x^2z +z^2x + \eps(q^2+ \eps( z+ x))}{x^2+ z^2+q^2 +2\eps(z-x+\eps)} \geq \frac{x^2z+z^2x}{x^2+z^2+y^2}=c_1.
	\end{equation}
	Finally, suppose that $z>\frac12$. The only change is that the bold $z$ in Eq.~\eqref{eq:exp_1_a} would change to $x+y$ (which is smaller than $z$). This only increases the expression and would thus not make it negative. 
	%
	%
	\item[Case~2:] $x\geq q > x-\eps$. As in Case~1, we may assume w.l.o.g. that $z\leq \frac12$ (and handle the complimentary case in the same way). Then $p_x \sim x^2, p_z \sim z^2$; and $p'_x\sim q^2, p'_z\sim (z+\eps)^2$.
	The difference from Case~1 is that $p'_x = q^2$ rather than $(x-\eps)^2$. Denote $\delta = x-q < \eps$, then 
		\begin{align*}
	c'_1 &= p'_x z+ p'_z x + p'_y \eps =  \frac{(x-\delta)^2 z+(z+\eps)^2x + y^2 \eps }{(x-\delta)^2+(z+\eps)^2+y^2}\\
	& = \frac{x^2z -2\delta xz +\delta^2 z +z^2x +2\eps zx + \eps^2 x+ y^2 \eps }{x^2-2\delta x +\delta^2 + z^2+2\eps z +\eps^2+y^2}	\\
	& = \frac{x^2z  +z^2x  + \eps(2zx + \eps x+ y^2 ) +\delta(-2xz +\delta z)}{x^2+ z^2+y^2+\eps(2 z +\eps)+\delta(-2x +\delta) }	\\
	& \geq \frac{x^2z  +z^2x  + \eps(2zx + \eps x+ q^2 ) +\delta(-2xz +\delta z)}{x^2+ z^2+q^2+\eps(2 z +\eps)+\delta(-2x +\delta) } \tag{as in case~1, w.l.o.g. $p_y=p'_y=q^2$}	
	\end{align*}
So the expression we need to lower bound is a bit more complicated. Note that the $\delta$ multiplier in the nominator is exactly the one in the denominator multiplied by $z$. The $\eps$ multiplier is multiplied by $\hat x:= \frac{2zx + \eps x+ q^2 }{2z + \eps}>x$. So we can rewrite $c'_1$ as
$$    c'_1= \frac{x^2z  +z^2x  + \eps \hat x \alpha_\eps -\delta z \beta_\delta}{x^2+ z^2+q^2+\eps\alpha_\eps-\delta\beta_\delta},$$
where $\alpha_\eps = 2z+\eps, \beta_\delta = 2z-\delta$.
It is easy to check that 	$\eps\cdot\alpha_\eps,\delta\cdot\beta_\delta$ are positive and monotonically increasing in their respective arguments $\eps,\delta$. 
We argue that
\labeq{xx}
{\frac{x^2z  +z^2x}{x^2+ z^2+q^2} \leq x.}
Indeed, divide both sides by $x$, then 
$$\frac{z^2+xz}{z^2+ x^2+q^2} \leq \frac{z^2+x^2+x(z-x)}{z^2+ x^2+q^2} \leq \frac{z^2+x^2+(z/2)^2}{z^2+ x^2+q^2} \leq 1,$$
where the last inequality is since $z\leq \frac12=2q$. 

We now further subdivide into 2 cases. The simple case is when $z\geq \hat x$. 
\begin{align*}
&z\geq \hat x >x \geq \frac{x^2z  +z^2x}{x^2+ z^2+q^2} &\Rightarrow \tag{by Eq.~\eqref{eq:xx}} \\
c'_1 &\geq   \frac{x^2z  +z^2x  + \eps \hat x \alpha_\eps -\delta z \beta_\delta}{x^2+ z^2+q^2+\eps\alpha_\eps-\delta\beta_\delta} \\
& \geq  \frac{x^2z  +z^2x  + \eps \hat x \alpha_\eps -\eps z \beta_\eps}{x^2+ z^2+q^2+\eps\alpha_\eps-\eps\beta_\eps} \\
& = \frac{x^2z +z^2x + \eps(q^2+ \eps( z+ x))}{x^2+ z^2+q^2 +2\eps(z-x+\eps)}  \geq c_1 \tag{by Eq.~\eqref{eq:c_1}}
\end{align*}

The uglier case is when $z<\hat x$. First, observe that 
$$z\leq \hat x = \frac{2zx + \eps x+ q^2 }{2z + \eps} = x+  \frac{q^2}{2z+\eps} \leq x+\frac{1}{32z},$$
that is, $x,z$ must be quite close to one another. Now, 
\begin{align*}
c'_1 &\geq   \frac{x^2z  +z^2x  + \eps \hat x \alpha_\eps -\delta z \beta_\delta}{x^2+ z^2+q^2+\eps\alpha_\eps-\delta\beta_\delta} \\
& > \frac{x^2z  +z^2x  + \eps  x \alpha_\eps -\delta z \beta_\delta}{x^2+ z^2+q^2+\eps\alpha_\eps-\delta\beta_\delta} \tag{$\hat x>x$}\\
=\frac{x^2z +z^2x + \eps(q^2+ \eps( z+ x))}{x^2+ z^2+q^2 +2\eps(z-x+\eps)}.
\end{align*}
We got an expression that is strictly smaller than the one we had in Eq.~\eqref{eq:c_1} so we cannot use our previous result from Case~1. However, we can feed it again into Wolfram Alpha, with the additional constraint $z\geq x+\frac{1}{32z}$ and verify that it is still always larger than $c_1$.

	\item[Case~3:] $q\geq x> x-\eps$. Then $p_x \sim q^2, p_y \sim y^2, p_z \sim z^2$; and $p'_x\sim q^2, p'_z\sim (z+\eps)^2, p'_y \sim y^2$. Also suppose first $y\leq q$. Thus $z\geq \frac12 \frac x+y$. This means that 
	$$c_1 = \frac{s_x (x+y) + s_z x}{s_x+s_y+s_z} = \frac{q^2(x+y) + z^2 x}{2q^2 + z^2},$$
	and
	\begin{align*}
	c'_1 &= \frac{q^2(x+y) + (z+\eps)^2 x + \eps q^2}{2q^2 + (z+\eps)^2} =  \frac{q^2(x+y) + z^2x +\eps(2zx +\eps x+ q^2)}{2q^2 + z^2 + \eps(2z+\eps)}
	\end{align*}
	Bounding the ratios of nominator and denominator factors:
	\begin{align*}
	&\frac{2zx +\eps x+ q^2}{2z+\eps} \geq \frac{2zx +\eps x}{2z+\eps} = x,\\
	&c_1 = \frac{q^2(x+y) + z^2 x}{2q^2 + z^2} \leq \frac{q^2(2x) + z^2 x}{2q^2 + z^2}=x.\\
	\end{align*}

	If $y\geq q$, then 
	$$c'_1 = \frac{q^2\min\{z,x+y\} + z^2x +\eps(2zx +\eps x+ y^2)}{q^2+y^2 + z^2 + \eps(2z+\eps)}$$
	
	\begin{align*}
	&\frac{2zx +\eps x+ y^2}{2z+\eps} = x+ \frac{y^2}{2z+\eps}\geq x+ \frac{y\frac{1}{4}}{z+(x+z)} \geq  x+\frac{y}{8},\\
	&c_1 \leq \frac{q^2(x+y) + z^2 x}{q^2 +y^2 + z^2} \leq \frac{q^2x + z^2 x}{q^2 +y^2 + z^2}+\frac{yq^2}{q^2 +y^2 + z^2}\\
	&\leq \frac{q^2x + z^2 x}{q^2 + z^2}+\frac{y \frac{1}{16}}{q^2 +y^2 + z^2}\leq x+y\frac{1}{16 (q^2)+2(\frac{1-q}{2})^2}\\
	&=x+y\frac{1}{16 \frac{4}{64}+2\frac{9}{64}} = x+\frac{y}{5.5} < x+\frac{y}{8}.\\
	\end{align*}
		Thus in either case, $c'_1\geq c_1$. 
\end{description}
\end{proof}

\begin{rproposition}{prop:QCD_approx}
The $\frac14$-QCD mechanism has an approximation ratio of $\frac76 \cong 1.166$, and this is tight.
\end{rproposition}

\begin{sproof}
Let $\vec a = (a_1,a_2,a_3)$ be a profile, and denote $x=d(a_1,a_2), y=d(a_2,a_3), z=d(a_1,a_3)$. We assume w.l.o.g.  $z\geq y\geq x$, thus the optimal point is $a_2$. The optimal social cost is $x+y$. 

We first argue that the approximation only becomes worse by moving $a_2$ to the mid point between  her neighbors. 
This means that we are left to find the worst instance among the instances with distances $(x,x,1-2x)$ for some $x\leq \frac13$. The optimum in such an instance is $2x$ whereas the social cost of $\frac14$-QCD can now also be expresses as a function of $x$ (the expression differs when $x<\frac13$ and $x>\frac13$). This means we can write the approximation ratio as a function of $x$, and get the lower bound by plugging in $x=\frac14$.  
\end{sproof}

\section{Beyond 3 agents}
We have several conjectures regarding a general number of agents on the circle. First, regarding the general version of the PCD mechanism.
\begin{conjecture}
The worst approximation ratio is obtained when all points are on the same semi-circle. 
\end{conjecture}

Next, we consider arbitrary mechanisms for $n$ agents. 

\begin{rconjecture}{conj:peaks_only}
For any $n$, the best strategyproof mechanism is peaks-only.
\end{rconjecture}

We can prove a somewhat weaker result: 

\begin{rproposition}{prop:antipod}
For any $n$, the optimal strategyproof mechanism w.l.o.g. only places the facility either on peaks, or on points antipodal to peaks. 
\end{rproposition}

\begin{proof}
In this proof we use the notation $f(\vec a)$ instead of $f_{\vec a}$, and use $p_f(x)$ as the probability that $f$ returns $x$ (when the profile is clear from context). 
For a profile $\vec a=(a_1,\ldots,a_n)$, denote by $b_i$ the point antipodal to $a_i$, and let $A=\{a_1,\ldots,a_n,b_1,\ldots,b_n\}$. 
Suppose that in some some profile $\vec a$, the mechanism $f$ places the facility with some probability $p$ on point $\alpha\notin A$. Denote by $\beta,\gamma$ the nearest points from $A$ clockwise and counterclockwise, respectively. Let $x=d(\alpha,\beta), y=d(\alpha,\gamma)$. 

We define a mechanism $f'$ that is identical to $f$, except that instead of setting $p_f(\alpha)=p$, it sets $p_{f'}(\alpha)=0$; $p_{f'}(\beta) = p_f(\beta) + p\frac{y}{x+y}$; and $p_{f'}(\gamma) = p_f(\gamma) + p\frac{x}{x+y}$ (we apply this at all profiles).

We claim that for any agent~$i$, $c_i(f(\vec a)) = c_i(f'(\vec a))$. This would show both that $f'$ is strategyproof (since $f$ is) and that $SC(f(\vec a),\vec a)=SC(f'(\vec a),\vec a)$ for all $\vec a$. 

Indeed, consider some agent placed at $a_i$. From the three points $\alpha,\beta,\gamma$, the one farthest from $a_i$ cannot be $\alpha$, since this would mean that $b_i$ (the point antipodal to $a_i$) is strictly in the open interval $(\beta,\gamma)$, whereas by construction there are no more points from $A$ in this interval. Thus w.l.o.g. $d(a_i,\beta)<d(a_i,\alpha)<d(a_i,\gamma)$, and thus 
$$ d(a_i,\beta) = d(a_i,\alpha)-d(\alpha,\beta) = d(a_i,\alpha)-x;~~ d(a_i,\gamma) = d(a_i,\alpha)+d(\alpha,\gamma) = d(a_i,\alpha)+y.$$
 We have that 
\begin{align*}
c_i(f'(\vec a),\vec a)  &=  p_{f'}(\beta) d(a_i,\beta)+ p_{f'}(\gamma) d(a_i,\gamma) + \sum_{t\in [m]\setminus \{\alpha,\beta,\gamma\}}p_{f'}(t) d(a_i,t)\\
&=   (p_f(\beta) + p\frac{y}{x+y}) d(a_i,\beta)+  ( p_f(\gamma) + p\frac{x}{x+y})d(a_i,\gamma) + \sum_{t\in [m]\setminus \{\alpha,\beta,\gamma\}}p_{f}(t) d(a_i,t)\\
&=   p\frac{y}{x+y} d(a_i,\beta)+  p\frac{x}{x+y}d(a_i,\gamma) + \sum_{t\in [m]\setminus \{\alpha\}}p_{f}(t) d(a_i,t)\\
&=   p\frac{y}{x+y}( d(a_i,\alpha)-x)+  p\frac{x}{x+y}( d(a_i,\alpha)+y) + \sum_{t\in [m]\setminus \{\alpha\}}p_{f}(t) d(a_i,t)\\
&=   p\frac{y}{x+y} d(a_i,\alpha)-p\frac{y}{x+y}x+  p\frac{x}{x+y} d(a_i,\alpha)+p\frac{x}{x+y}y + \sum_{t\in [m]\setminus \{\alpha\}}p_{f}(t) d(a_i,t)\\
&=   p\frac{y}{x+y} d(a_i,\alpha)+  p\frac{x}{x+y} d(a_i,\alpha) + \sum_{t\in [m]\setminus \{\alpha\}}p_{f}(t) d(a_i,t)\\
& =  p d(a_i,\alpha) + \sum_{t\in [m]\setminus \{\alpha\}}p_{f}(t) d(a_i,t) = \sum_{t\in [m]}p_{f}(t) d(a_i,t) = c_i(f(\vec a),\vec a),
\end{align*}
as required.
\end{proof}
We remark that such a proof would not work for Conjecture~\ref{conj:peaks_only}, since it is possible to construct mechanisms that use an antipodal point to balance incentives and maintain strategyproofness. Our conjecture is thus that this can only improve the social cost when the social cost is far from being optimal.

\section{General Graphs}

\begin{rtheorem}{thm:PD_SP}
The PD mechanism is strategyproof in expectation for 3 agents in any metric space (in particular on any graph).
\end{rtheorem}
\begin{figure}
%
\tikzstyle{VertexStyle}=[circle,draw=black,fill=white,
inner sep=0pt,minimum size=4mm]
\centering
\tikzset{
  font={\fontsize{10pt}{10}\selectfont}}
\begin{tikzpicture}[scale=1,transform shape]
  \Vertex[x=0,y=0,L=$$]{v00}
	\Vertex[x=1,y=0,L=$$]{v10}
\Vertex[x=2,y=0,L=$$]{v20}
\Vertex[x=3,y=0,L=$$]{v30}
\Vertex[x=4,y=0,L=$a'_1$]{v40}
\Vertex[x=5,y=0,L=$$]{v50}
\Vertex[x=0,y=1,L=$a_3$]{v01}
\Vertex[x=1,y=1,L=$$]{v11}
\Vertex[x=2,y=1,L=$$]{v21}
\Vertex[x=3,y=1,L=$$]{v31}
\Vertex[x=4,y=1,L=$a_1$]{v41}
\Vertex[x=5,y=1,L=$$]{v51}
\Vertex[x=0,y=2,L=$$]{v02}
\Vertex[x=1,y=2,L=$$]{v12}
\Vertex[x=2,y=2,L=$$]{v22}
\Vertex[x=3,y=2,L=$a_2$]{v32}
\Vertex[x=4,y=2,L=$$]{v42}
\Vertex[x=5,y=2,L=$$]{v52}
  
	\Edge(v00)(v10)
	\Edge(v10)(v20)
	\Edge(v20)(v30)
	\Edge(v30)(v40)
	\Edge(v40)(v50)
	\Edge(v50)(v51)
	\Edge(v51)(v41)
	\Edge(v41)(v31)
	\Edge(v31)(v21)
	\Edge(v21)(v12)
	\Edge(v12)(v11)
	\Edge(v11)(v10)
	\Edge(v01)(v11)
	\Edge(v01)(v02)
	\Edge(v12)(v22)
	\Edge(v22)(v32)
	\Edge(v00)(v01)
	\Edge(v41)(v52)
	\Edge(v52)(v42)
	%
%
%
%

\end{tikzpicture}
%

\caption{\label{fig:graph}Example of a profile and a possible manipulation on a graph. All edges have the length $1$. The distances in profile $\vec a$ (before normalization) are $x=d(a_1,a_2) = 5, y=d(a_2,a_3) = 4, z=d(a_1,a_3)=5$. In the modified profile $\vec a'=(a'_1,a_2,a_3)$, the distances are $x'=7,z'=5$, and $\eps = 3,\alpha=-2,\beta=0$. }
\end{figure}
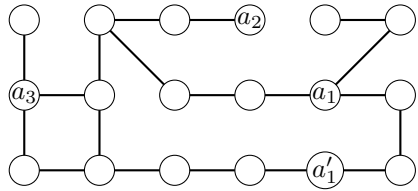

\begin{proof} Let $f=f^{PD}$ be the Proportional Distance mechanism.
Consider a deviation $a'_1$ by agent~1 that results in the profile $\vec a'=(a'_1,a_2,a_3)$, and denote the new distances $x'=d(a'_1,a_2)$ and $z'=d(a'_1,a_3)$. We normalize distances such that $y=1$. Denote $\eps = d(a_1,a'_1)$, $\alpha = x'-x$ and $\beta = z'-z$ (see Fig.~\ref{fig:graph} for an example). By triangle inequality:
\begin{align}
&|\alpha|,|\beta|\leq \eps\\
&|x-z|\leq y =1
\end{align} 

The cost of truthful reporting by agent~1 is 
$$c_1(f(\vec a)) = f_{\vec a}(a_1)0+f_{\vec a}(a_2)x + f_{\vec a}(a_3)z = \frac{2xz}{1+x+z}.$$

On the other hand, after reporting $a'_1$, the cost is
$$c_1(f(\vec a')) = f_{\vec a'}(a'_1)\eps+f_{\vec a'}(a_2)x + f_{\vec a'}(a_3)z = \frac{\eps+ x'z + xz'}{1+x'+z'} =  \frac{\eps+ (x+\alpha) z + x(z+\beta)}{1+x+z+\alpha+\beta}.$$

We need to show that $c'_1-c_1 \geq 0$. We begin as follows.
$$c_1(f(\vec a'))-c_1(f(\vec a)) = \frac{(\eps+ (x+\alpha) z + x(z+\beta))(1+x+z) - 2xz(1+x+z+\alpha+\beta)}{(1+x+z)(1+x+z+\alpha+\beta)}.$$
Since the denominator is positive, we focus on the nominator. W.l.o.g. $x\geq z$. Denote $\delta = x-z$ and note that $\delta\in [0,1]$. 
\begin{align*}
c_1(f(\vec a'))-c_1(f(\vec a)) &=_{sign} (\eps+ (x+\alpha) z + x(z+\beta))(1+x+z) - 2xz(1+x+z+\alpha+\beta) \\
&=\eps(1+x+z)+ 2xz(1+x+z) + \alpha z +\alpha zx +\alpha z^2 + \beta x + \beta zx + \beta x^2 \\
&~~~~~~~~~~- 2xz(1+x+z) -2\alpha xz-2\beta xz \\
&= \eps(1+x+z) + \alpha z -\alpha zx +\alpha z^2 + \beta x - \beta zx + \beta x^2\\
& = \eps(1+x+z)  + \alpha z (1+z-x) + \beta x (1+x-z)  \\
& = \eps(1+2z+\delta)  + \alpha z (1-\delta) + \beta (z+\delta)(1+\delta)
\end{align*}
Note that $z(1-\delta)$ and $(z+\delta)(1+\delta)$ are nonnegative. Thus we can lower bound the expression by taking the lower bound of $\alpha$ and $\beta$, which is $-\eps$. Therefore
\begin{align*}
c_1(f(\vec a'))-c_1(f(\vec a)) &=_{sign} \eps(1+2z+\delta)  + \alpha z (1-\delta) + \beta (z+\delta)(1+\delta)\\
& \geq  \eps(1+2z+\delta)  + (-\eps) z (1-\delta) + (-\eps) (z+\delta)(1+\delta)\\
& = \eps ( 1+ 2z + \delta -z +z\delta -z -z\delta -\delta -\delta^2)  = \eps ( 1  -\delta^2) \geq 0,
\end{align*} 
as required.\QED  
\end{proof}

\begin{proposition}The PD mechanism dominates the RD mechanism. 
\end{proposition}

\begin{proof}
W.l.o.g. $x\leq y \leq z$ and $x+y+z=1$. Denote $\alpha=x-\frac13, \beta=y-\frac13, \gamma=z-\frac13$, then $\alpha+\beta+\gamma=0$ and $\alpha\leq 0 \leq \gamma$.
\begin{align*}
SC(f^{PD}(\vec a)) &= x(y+z) + y(x+z) + z(x+y) = (\frac13+\alpha)(y+z) + (\frac13+\beta)(x+z) + (\frac13+\gamma)(x+y)\\
&= SC(f^{RD}(\vec a)) +   (\alpha+\beta)z + (\alpha+\gamma)y +  (\beta+\gamma)x\\
&=SC(f^{RD}(\vec a))   -\gamma z + (\alpha+\gamma)y -\alpha x =  SC(f^{RD}(\vec a)) +   \gamma (y-z) + \alpha(y-x) \leq SC(f^{RD}(\vec a)),
\end{align*}
Since $\gamma \geq 0 \geq y-z$ and $\alpha \leq 0 \leq y-x$. The inequality is strict for any profile where $x<y<z$. \QED
\end{proof}

%
%
%

\section{Euclidean Spaces}\label{sec:plane}
We conclude the technical part of the paper with a contribution to the analysis of \emph{deterministic} mechanisms in Euclidean spaces, focusing mainly on 3 agents in the plane.

Suppose that $\calX$ is a convex subset of an Euclidean space $\mathbb R^D$ with the $\ell_2$ norm. Every agent location $a_i$ is a vector $(a_{ij})_{j\leq D}$.

\paragraph{The Multi-Median mechanism}
Consider the (deterministic) \emph{multi-median (MM) mechanism}, which takes the median independently on each dimension. That is, $f^{MM}(\vec a) = (x_1,\ldots,x_D)$, where each $x_j$ is the median of $(a_{ij})_{i\leq n}$ (breaking ties according to some fixed order).

The MM mechanism is strategyproof for the same reason that the (one-dimensional) median is: an agent can only move the facility away from her own  location in each dimension.\footnote{Note that in contrast to the one-dimensional median, the MM mechanism is not group-strategyproof, as each agent may agree to suffer a small loss in one dimension to gain more in another dimension.}

\begin{proposition}\label{prop:MM_D}The MM mechanism  has an approximation ratio of at most $\sqrt D$ for any number of agents $n$.
\end{proposition}
\begin{proof}[Proof sketch]
The proposition follows from the fact that the MM mechanism is optimal for the $\ell_1$ norm, and that switching from  $\ell_1$  to $\ell_2$ may change the norm of a vector by a factor of at most $\sqrt D$. 
\end{proof}

\subsection{The 2-dimensional Plane}
\begin{remark}
While I was working on this problem for three agents, Goel and Hann-Caruthers~\shortcite{goel2019coordinate}  solved the problem for any number of agents on the plane. They showed that the tight approximation  bound of the MM mechanism is $\frac{\sqrt 2 \sqrt{n^2+1}}{n+1}$, which entails $\sqrt{\frac54}\cong 1.118$ for $n=3$ agents.  I still report my findings as they apply different techniques that may be of interest. 
\end{remark}

 The point that minimizes the sum of distances to vertices of a triangle is known as the Fermat (or Fermat-Torricelli) point. We use two known characterizations of it in this section: a geometric characterization to prove a lower bound, and an algebraic characterization for an upper bound.

\begin{figure}
\begin{tikzpicture}[scale=0.8,transform shape]
\tikzset{
  font={\fontsize{12pt}{12}\selectfont}}
	
	\draw[dashed] (3,1.73) -- (6,3.46);
	\draw[dashed] (6,0) -- (6,3.46);
	\node[fill=white] at (6,3.46) {$D$};

	\draw[dotted,thick] (5,1.73)+(180:2) arc (180:300:2);
		\node at (5,1.73) {$Q$};
		\node at (4.5,0.6) {$Z$};
	\draw[ultra thin] (5*0.98,1.73*0.98) -- (0,0);
  \Vertex[x=0,y=0,L=$C$]{C}
  \Vertex[x=3,y=1.73,L=$B$]{B}
	 \Vertex[x=6,y=0,L=$A$]{A}
	\draw[|<->|] (0,-.6) -- (2.98,-.6);
	\node at (1.5,-1) {$1$}; 
		\draw[|<->|] (3.02,-.6) -- (6,-.6);
	\node at (4.5,-1) {$1$}; 

\draw[|<->|] (6.5,0.02) -- (6.5,1.73);
	\node at (6.8,1.25) {$\frac{1}{\sqrt 3}$};
	
	  \tikzstyle{LabelStyle}=[fill=white,sloped,above]
	\Edge[](B)(C)
		\Edge[](A)(C)
			\Edge[](A)(B)
			
			\node[fill=white] at (3,0) {$\times$};
			

\end{tikzpicture}
\caption{\label{fig:tri2}A profile where an agent located in $C$ has a manipulation in the optimal mechanism. This is also the profile where the multi-median obtains its worst approximation ratio in Case~Ib ($\times$ marks the output of the MM mechanism).
}
\end{figure}
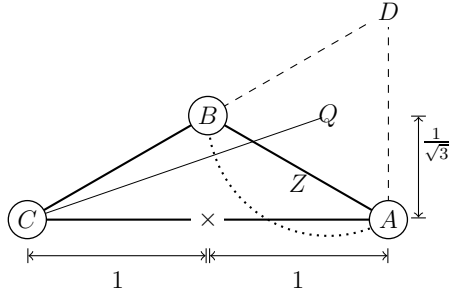
It is known that MM is the only deterministic strategyproof mechanism that is also Pareto optimal and anonymous, and this is true for any odd number of agent in the plane~\cite{peters1992pareto}. Since the optimal mechanism is in particular Pareto optimal and anonymous, it cannot be strategyproof. Yet, we are not aware of an explicit lower bound on the approximation ratio of (deterministic) strategyproof mechanisms.
\begin{proposition}\label{prop:n3_plane_lower}
The optimal mechanism for 3 agents in the plane is not strategyproof. Further, any deterministic strategyproof mechanism has an approximation ratio strictly above $1.0002$.
\end{proposition}
\begin{proof}[Proof sketch]
Our proof relies on a known geometric characterization of the Fermat point $\vec z^*$ in a triangle $ABC$:\footnote{See solution~6 here: \url{http://www.cut-the-knot.org/Generalization/fermat_point.shtml}}  
\begin{itemize}
	\item In triangles with some obtuse angle of $120^\circ$ or more, $\vec z^*$ is simply the vertex on the obtuse angle, and thus $SC(\vec z^*)$ is the sum of the two shorter edges. In other triangles, the Fermat point can be found as follows.
	\item  Consider some edge of the original triangle $ABC$ (say $AB$), the equilateral triangle $ABD$ built on the other side of $AB$, and its escribed circle whose center is the point $Q$. Consider the arc of the circle that intersects $AB$ (the dotted arc in Fig.~\ref{fig:tri2}), and denote by $Z$ the area enclosed by the arc and the edge $AB$. 
\item  If $C$ is outside the area $Z$, then the Fermat point is the intersection of the line $CD$ with the dotted arc. If $C$ is inside $Z$, then $C$ forms an angle of at least $120^\circ$ and becomes the Fermat point itself. 
\end{itemize}

Now, consider a profile where $a_1=A,a_2=B$. Agent~3 can set the Fermat point to be anywhere in the slice $Z$. When $a_3=C$, as in the figure, then the Fermat point is $B$. However, clearly $B$ is \emph{not} the closest point to $a_3=C$ in the slice $Z$. The closest point is the intersection $I$ of the thin line $CQ$ with the dotted arc, so reporting $a'_3=I$ would also move the Fermat point to $I$, and thus be a manipulation for agent~3. 

We can further extend the proof above to give an explicit lower bound (of $1.0002$) on the approximation ratio of any deterministic strategyproof mechanism: to get a better approximation ratio, the selected point in profile $ABC$ must be close to the Fermat point $B$, and in profile $ABI$ must be close to the Fermat point $I$. For an approximation of $1.0002$ or better, the location of the facility in profiles $ABC$ and $ABI$ must be disjoint.  However then moving agent~3 from $C$ to $I$ would still be a manipulation, as shown above. 
\end{proof}
 
\rmr{
This means that an agent placed on $C$ can essentially move the Fermat point to any point $q\in J$ (e.g. by placing herself directly on $q$). 

In particular, suppose that $ABC$ is isosceles with $B$ being a $120^\circ$ angle. Then $B\in J$ is also the Fermat point (note that $ABD$ form a straight line). However, the closest point to $C$ on $J$ (called $P$) is obtained in the intersection of $J$ and the line $CQ$, not $CD$. Thus the agent on $C$ has a manipulation by moving to point $P$. 

To get a lower bound on the approximation, suppose that $|AB|=2\sqrt 3$. To get an approximation of $\leq 1.0002$, the facility $B'$ in profile $ABC$ must be within a small distance (at most $0.1$) from the Fermat point $B$ (in addition, we can pick the agent in the far side of $B'$, so that $d(B',C)\geq d(B,C)-0.05$). Similarly, in the profile $ABP$ the facility $P'$ must be close (distance at most $0.1$) from the Fermat point $P$. However, we then get that 
$$d(C,B')\geq d(C,B)-0.05 \geq 3.464-0.05 > 3.4 > 3.292 + 0.1 \geq d(C,P) + 0.1 \geq d(C,P'),$$
and thus moving from $C$ to $P$ (an moving the facility from $B'$ to $P'$) is a manipulation. 

}

This raises the question of the best approximation ratio that can be guaranteed, even with deterministic strategyproof mechanisms. 
In the plane $D=2$, so Prop.~\ref{prop:MM_D} provides us with an approximation ratio of $\sqrt 2\cong 1.41$, which is somewhat better than the approximation of RD ($2-\frac2n$)---but only for $n\geq 4$. We still need to show that the multi-median beats $\frac43$ for $n=3$ as well.
\begin{theorem}\label{thm:n3_plane}For $n=3$ agents in the plane, the multi-median mechanism has a worst-case approximation ratio of $\sqrt{\frac{5}{4}} \cong 1.118$, and this bound is tight.  
\end{theorem}
\begin{proof}
Note that the MM mechanism is invariant to translation, scaling, rotation at right angles, and mirroring of agents' locations. Thus given any profile $(a_1,a_2,a_3)$, w.l.o.g. the order along the horizontal axis is $a_3,a_2,a_1$ (with $a_2$ strictly to the right of $a_3$) and along the vertical axis $a_2$ is weakly above $a_3$. We the translate $a_3$ to $C=(0,0)$, and scale everything such that $a_2$ is mapped to $B=(1,x)$ for some $x\geq 0$. The last point $a_1$ is mapped to some $A=(1+y,-z)$ where $y\geq 0$. 

We can thus describe any profile $\vec a\in \calX^3$ (up to translation, scaling, and mirroring) using the three parameters $x,y,z$. We refer to the case where $z> 0$ as Case~I, and to $z\leq 0$ as Case~II. 

We denote $a=d(BC), b=d(AC)$ and $c=d(AB)$.
See Fig.~\ref{fig:tri} for an example. 
\begin{figure}

\tikzstyle{VertexStyle}=[circle, thin,draw=black,fill=white,
inner sep=0pt,minimum size=8mm]
\centering

\tikzstyle{dot}=[circle,draw=black,fill=white,
inner sep=0pt,minimum size=4mm]
\tikzset{
  font={\fontsize{10pt}{10}\selectfont}}
\label{sfig:tri2}
\begin{tikzpicture}[scale=0.8,transform shape]

\draw[dashed,thin] (0,0) -- (4,0);
\draw[dashed,thin] (4,2.5) -- (4,0);
  \Vertex[x=0,y=0,L=$C$]{C}
  \Vertex[x=4,y=2.5,L=$B$]{B}
	 \Vertex[x=6,y=-2,L=$A$]{A}
	\draw[|<->|] (0,3) -- (3.98,3);
	\node at (2,3.3) {$1$}; 
		\draw[|<->|] (4.02,3) -- (6,3);
	\node at (5,3.3) {$y$}; 

\draw[|<->|] (6.5,0.02) -- (6.5,2.5);
	\node at (6.8,1.25) {$x$};
	
	\draw[|<->|] (6.5,-0.02) -- (6.5,-2);
	\node at (6.8,-1) {$z$};
	  \tikzstyle{LabelStyle}=[fill=white,sloped,above]
	\Edge[label=$a$](B)(C)
		\Edge[label=$b$](A)(C)
			\Edge[label=$c$](A)(B)

			\node at (4,0) {$\times$};
	\node at (3,1) {$*$};
	
				\draw (0,0)+(-18:0.8) arc (-18:32:0.8);
			\node at (1.1,-0.2) {$\gamma$};
		\node at (1.4,0.3) {$90^\circ-\alpha$};
	
				\draw (4,2.5)+(-66:0.8) arc (-66:-148:0.8);
			\node at (3.6,1.6) {$\alpha$};
		\node at (4.2,1.5) {$\beta$};
	
\end{tikzpicture}
%
%
%
\caption{\label{fig:tri}Above,  the triangle defined by $x=\frac58,y=\frac12,z=\frac12$. The output of the MM mechanism $\vec x^*$ and the Fermat-Torricelli point $\vec z^*$ are marked by $\times$ and $*$, respectively. 
}
\end{figure}
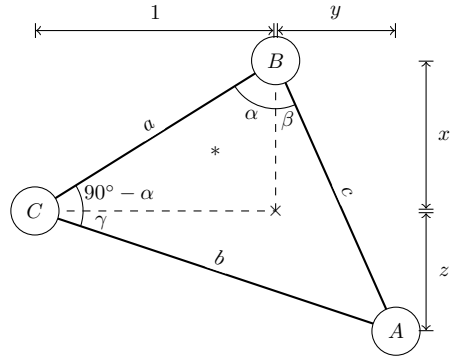

\begin{table*}[t]
\begin{center}
\begin{tabular}{|l||c|c|c|c|}
\hline
      metric space  & Any & Plane &  Circle & Tree/Line \\
	\hline
UB & $2$ (dictator)  & $1.118$ (MM, Thm.~\ref{thm:n3_plane}) & $2$ (dictator) &$1$ (median) \\
LB & $2$ (*)  &  $>1$ (Prop.~\ref{prop:n3_plane_lower})     & $2$ (*) & $1$ \\
\hline
\end{tabular}
\caption{\label{tab:results_d}A summary of approximation bounds for 3-agent deterministic mechanisms. (*) - from \cite{SV04,DFMN:2012:EC}. \vspace{-4mm} }
\end{center}
\end{table*}

\newpar{Case~I}
The lengths of the three edges are 
$$(i)~~ a=\sqrt{1+x^2}; (ii)~~ b=\sqrt{y^2+(z+x)^2}; (iii)~~  c=\sqrt{z^2+(1+y)^2}.$$
By the way we constructed the triangle, $B$ provides the median on axis~1, and $C$ provides the median on axis~2, thus the multi-median is the point $\vec x^*=(1,0)$.

It is easy to compute $SC(\vec x^*) = d(A,\vec x^*)+d(B,\vec x^*)+d(C,\vec x^*) =  \sqrt{y^2+z^2}+x+1$.
Recall that in triangles with some obtuse angle of $120^\circ$ or more (case~Ia), $\vec z^*$ is simply the vertex on the obtuse angle, and thus $SC(\vec z^*)$ is the sum of the two shorter edges. 
For Case~1a, if we assume symmetry, we get (by taking derivative w.r.t. $x$) that  the worst case for the isosceles triangle is $x=0.5,y=1,z=0$ (in particular the obtuse angle is strictly more than $120^\circ$).\footnote{I thank Wade Hann-Caruthers for helping pointing out a mistake in the original analysis.}
 Then $SC(\vec x^*)=2.5$ and $SC(\vec z^*)=2\sqrt{1.25}$, so we get an approximation ratio of
$$\frac{2.5}{2\sqrt{1.25}} = \frac{2.5}{\sqrt{5}} = \frac{2.5\sqrt 5}{5} = \frac{2\sqrt 5}{4}=\sqrt{\frac{20}{16}} \cong 1.118.$$ 

It is easy to verify that breaking the symmetry of the triangle will only improve the approximation ratio. 

In other triangles (case~Ib), we get the following unattractive expression based on an algebraic characterization of $\vec z^*$:\footnote{We could not find this explicit expression in published literature, however a  short proof is given by Quang Hoang in  StackExchange~\cite{Hoang_MO}.}
\begin{small}
$$SC(\vec z^*)^2 = \frac12\(a^2+b^2+c^2 + \sqrt{3(a+b+c)(-a+b+c)(a-b+c)(a+b-c)}\).$$
\end{small}

We solve this case using computer optimization. 
We need to constrain the values of $x,y,z$ so as to avoid angles larger than $120^\circ$. Denote $\alpha \defeq \angle CB\vec x^*, \beta \defeq \angle AB\vec x^*, \gamma \defeq \angle AC\vec x^*$ (see Fig.~\ref{fig:tri}). We have that
$$(iv)~~ \tan \alpha = \frac1x; (v)~~ \tan \beta = \frac{y}{x+z}; (vi)~~ \tan \gamma = \frac{z}{1+y}.$$
Angle $\angle CAB$ can never be obtuse, so we only need the constraints on angles $\angle ABC$ and $\angle ACB$:
$$(I)~~ \alpha+\beta  \leq 120^\circ;  (II)~~ 90^\circ-\alpha + \gamma \leq 120^\circ.$$
To find the worst instance, we need to maximize
\begin{scriptsize}
$$\frac{SC(\vec x^*)}{SC(\vec z^*)} = \frac{ \sqrt{y^2+z^2}+x+1}{\sqrt{\frac12\(a^2+b^2+c^2 + \sqrt{3(a+b+c)(-a+b+c)(a-b+c)(a+b-c)}\)}},$$
\end{scriptsize}
subject to the equalities $(i)-(vi)$, inequalities $(I),(II)$, and non-negativity of $x,y,z$. Since $a$ and $b$ are bounded from $0$, all derivatives in $x,y$ and $z$ are bounded. We used  grid search to verify that the maximum is  obtained at $x=\frac{1}{\sqrt 3},y=1,z=0$. This is an isosceles triangle with a sharper angle than in Case~Ia (exactly $120^\circ$) so the approximation ratio is strictly better. 

\newpar{Case~II} If $z$ is negative, we rotate the triangle $90^\circ$ counterclockwise, and scale down by a factor of $x$. Then we are back at case~$I$, where the vertices switch roles. 
\end{proof}


 Note that by using randomization, we are likely to get a certain improvement. For example, we can select the axes according to a random rotation, and then run the multi-median mechanism.
We leave the analysis of such mechanisms to future work. It is an open question e.g. whether some ``random rotation multi-median'' has a constant approximation ratio in high dimensions.

\end{document}